\documentstyle[12pt]{JHEP3}
\newcommand{\bej}[1]{ \begin{equation}\label{#1} }
\newcommand{\eej}{\end{equation}}
\newcommand{\beaj}[1]{\begin{eqnarray}\label{#1} }
\newcommand{\eeaj}{\end{eqnarray}}
\newcommand{\eq}[1]{(\ref{#1})}
\def\ZZZ{{\hskip-3pt\hbox{ Z\kern-1.6mm Z}}}
\def\zzz{{\hskip-3pt\hbox{ z\kern-1mm z}}}

\newcommand{\be}{\begin{equation}}
\newcommand{\ee}{\end{equation}}
\newcommand{\ben}{\begin{eqnarray}\displaystyle}
\newcommand{\een}{\end{eqnarray}}
\newcommand{\refb}[1]{(\ref{#1})}
\newcommand{\p}{\partial}

\def\one{{\hbox{ 1\kern-.8mm l}}}
\def\zero{{\hbox{ 0\kern-1.5mm 0}}}

\title{S-matrix for magnons in the D1-D5 system}
\author{Justin R. David {${}^a$}
and Bindusar Sahoo{${}^b$} \\
${}^a$ Centre for High Energy Physics,\\
Indian Institute of Science,\\ C.V. Raman Avenue, Bangalore 560012, India. \\
\email{justin@cts.iisc.ernet.in}\\
${}^b$  High Energy Section, \\
International Centre for Theoretical Physics, \\
Strada Costeria, Trieste, Italy.   \\
\email{bsahoo@ictp.it}
}

\abstract{
We show that integrability and symmetries of the near horizon geometry of the
D1-D5 system  determine the S-matrix for the scattering of magnons 
with polarizations in $AdS_3\times S^3$  completely  up to a phase. Using semi-classical
methods we evaluate the phase to the leading and to the one-loop
approximation in the strong coupling expansion.
We then show that the  phase obeys the
 unitarity constraint implied by  the  crossing relations to the one-loop order.
We also verify
that the dispersion relation obeyed by these magnons is one-loop exact at
strong coupling which is  consistent with their BPS nature.
}
\preprint{IC/2010/014}

\begin{document}

\section{Introduction}

Integrability has played an important role in understanding the spectrum
of ${\cal N}=4$ super Yang-Mills in the planar limit.  For this system
integrable structures  were discovered both in the field theory and
for strings propagating on $AdS_5\times S^5$, its holographic dual.
In fact the integrable structures found on both sides of the
holographic duals played complementary roles in discovering the
exact spectrum of planar  ${\cal N}=4$ super Yang-Mills.
For a recent review and a complete list of references see
\cite{Serban:2010sr}.
Another well studied dual pair is the duality between the
IIB string theory on $AdS_3\times S^3\times {\cal M}$ where
${\cal M}$ could be $T^4$ or $K^3$ and the ${\cal N}=(4,4)$
superconformal field theory on a resolution of the symmetric
product \cite{Maldacena:1998bw}
\begin{equation}
\label{sym-prod}
{\cal M}^N/S(N).
\end{equation}
In this paper we will restrict our attention to the case when ${\cal M }$ is
$T^4$.
$AdS_3\times S^3\times T^4$ arises as a near horizon limit of
the system of $Q_1$ D1-branes and $Q_5$ D5-branes wrapped on
$T^4$, then $N$ in (\ref{sym-prod}) is given by
$Q_1Q_5$.  String theory on $AdS_3\times S^3$ is known
to be classically  integrable \cite{Bena:2003wd}.
There has been recent progress in organizing conformal perturbation theory
on the symmetric product \cite{Pakman:2009ab,Pakman:2009zz,Pakman:2009mi}
 however similar integrable structures
have not yet been discovered for the ${\cal N}=(4,4)$ on the symmetric product.

One  direction to explore the role of integrability in this system is to
precisely determine the implication of integrability on its spectrum.
Recently \cite{Babichenko:2009dk} has put forward
 a proposal for the quantum Bethe equations from examining classical
Bethe equations of the string theory on $AdS_3\times S^3$.
 In this paper we will
study the implication of integrability on a certain class of excitations
which are present both in the symmetric product as well as on
the string theory on $AdS_3\times S^3\times T^4$.
These excitations are similar to the magnon excitations in
${\cal N}=4$ Yang-Mills which played a crucial role in understanding
integrability for this case.
For the D1-D5 system they were studied in \cite{David:2008yk} and were argued to be
BPS in a centrally extended $SU(1|1)\times SU(1|1)$ algebra. The
BPS condition then determined
their dispersion relation.  Briefly they are states of the form
\begin{equation}
J_{p_1}^{-1} J_{p_2}^{-1}  \cdots J_{p_j}^{-1} |0\rangle \otimes |0\rangle,
\end{equation}
where the vacuum denotes $\ZZZ_J$ twisted chiral primary of the
${\cal N}=(4,4)$ conformal field theory with R-charge $(\frac{J-1}{2}, \frac{J-1}{2})$.
$J_{p}^{-1}$ are operators which lower the left $J^3$ quantum number and carry momentum $p$ in the $\ZZZ_J$ twisted sector.
 Under the action of an element of $\ZZZ_J$, $J_p^{-1} |0\rangle \otimes |0\rangle$
 picks up a phase proportional to integer multiples of $p$.
 These states were argued to be BPS and as a result the dispersion relation
  for a single magnon is given by
 \begin{equation}
 \label{intro-1}
 \Delta -J = \sqrt{1 + 16 g^2 \sin^2\frac{p}{2} },
 \end{equation}
 where $\Delta$ is the sum of the
 left and right moving  conformal dimension of the state and $J$ is the
 sum of the left and right moving $J^3$ charge.
 $g$ is a function of the parameters of the D1-D5 system.
 A similar dispersion relation can be written for a bound state of
$Q$ magnons, the dyonic magnon which is given by 
\begin{equation}
\label{intro-2}
 \Delta -J = \sqrt{Q + 16 g^2 \sin^2\frac{p}{2} }.
 \end{equation}
 Both at strong coupling that is
 at the zeroth order in the
 string sigma model and at first order in conformal perturbation theory
 on the symmetric product
 it was found that
 \begin{equation}
 \label{ex-rel}
 16 g^2 = \frac{g_6^2  Q_1Q_5}{\pi^2}.
 \end{equation}
 Here $g_6$ is the 6-dimensional string coupling.

In this paper we use the $SU(1|1)$  symmetry
and integrability to constrain the
S-matrix for the scattering of these magnons up to a phase. Note that these magnons
have polarizations within the $AdS_3\times S^3$. 
Since we rely  purely on symmetry this argument can be applied both
to the string side and the conformal field theory on the symmetric product.
The phase cannot be determined by symmetry considerations alone.
To determine the phase
we  restrict ourselves to the $SU(2)$ sector and
examine the  dyonic magnons as classical solutions of the string sigma
model.
We then use the semi-classical string theory to determine this phase to
leading and to one loop order in the string sigma model.
To determine the one loop correction we use the method developed by
\cite{Chen:2007vs} for the case of magnons in $AdS_5\times S^5$.
This method is based  on
determining the phase shifts suffered by plane
waves scattering off the magnon.
We then verify that this phase obeys the unitarity constraint implied by the
crossing symmetry equations.
We also determine corrections to
the  dispersion (\ref{intro-1}), (\ref{intro-2})
relations at one loop in strong coupling and show that
there is no corrections to the relation (\ref{ex-rel}) at one loop in the string
sigma model.

The paper is organized as follows: The next section introduces the magnon
states of the D1-D5 system both in the symmetric product and
in the sigma model and
discusses their dispersion relation. In section 3 we use $SU(1|1)$ symmetry
and integrability to constrain the S-matrix for the scattering of these magnons
up to a phase. We also determine the phase at the zeroth order in the
string sigma model.
In section 4 we determine the one loop correction
to both the phase and the dispersion relation.  In section 5 we verify that
the phase determined to the one loop order satisfies the unitarity constraint
implied by the crossing symmetry equations. We end with some brief conclusions.
Appendix A provides the details regarding the derivation of the dispersion
relation of magnons to make the paper self contained. Appendix B discusses the
dressing method to obtain phase shifts. Appendix C derives the crossing
relations for the S-matrix with $SU(1|1)$ symmetry using the
antipode operation.

\section{Magnons in the D1-D5 system}

In this section we introduce the magnon excitations of the
D1-D5 system both in the boundary theory as well as  its semi-classical description as
classical solutions of the sigma model on $AdS_3\times S^3\times T^4$.
This section will also serve to set our notations and conventions.

\subsection{Magnons in the symmetric product}

The boundary theory corresponding to the system of $Q_1$ number of D1-branes
and $Q_5$ number of D5-branes in type IIB on $T^4$ is given by the
${\cal N}=(4, 4)$ superconformal field theory on a resolution of the
symmetric product orbifold \footnote{See \cite{David:2002wn} for a review.}.
\begin{equation}
\label{orbifold}
 {\cal M}= ( T^4)^{Q_1Q_5}/S(Q_1 Q_5).
\end{equation}
The global part of the ${\cal N}=(4,4)$ algebra is given by the supergroup
$SU(1,1|2)\times SU(1, 1|2)$. The two copies arise from the left movers and the right
movers of the conformal field theory on ${\cal M}$. The bosonic part of the supergroup
$SU(1,1|2)$ consists of the global part of the conformal algebra $SL(2, R)$ whose
generators are $L_0, L_{\pm}$ and the global part of the R-symmetry group $SU(2)_R$ whose generators are $J^3, J^\pm$. The $8$ supercharges of
$SU(1,1|2)$ are labeled by:
\begin{equation}
 G^{ab}_{1/2}, \qquad\hbox{and} \qquad G^{ab}_{-1/2},
\end{equation}
where $a \in \{+, -\}$ denotes the quantum numbers of the charges under $SU(2)_R$ and
$b \in \{+, -\}$ denotes the quantum numbers of charges under $SU(2)_I$ which is an
outer automorphism of the ${\cal N}=(4,4)$ algebra. The subscripts $+1/2$
denotes a weight of $L_0=-1/2$, while
a $-1/2$ denotes a weight of $L_0 =1/2$.
From the ${\cal N}=(4,4)$ algebra it is easy to see that the
set of generators
\begin{equation}
\label{subgroup-1}
\{ G_{-1/2}^{++} ,  G_{1/2}^{--}, L_0 -  J^3\} \quad\hbox{or} \quad
\{ G_{1/2}^{-+}, G_{-1/2}^{+-}, L_0-  J^3 \},
\end{equation}
each form a $SU(1|1)$ sub-algebra with the common central charge $L_0- J^3$.
From the right moving copy of $SU(1,1|2)$, we see  that the set of generators
\begin{equation}
\label{subgroup-2}
 \{ \tilde G_{-1/2}^{++} ,  \tilde G_{1/2}^{--}, \tilde L_0 -  \tilde J^3\} \quad\hbox{or} \quad
\{ \tilde G_{1/2}^{-+}, \tilde G_{-1/2}^{+-}, \tilde L_0 - \tilde J^3 \},
\end{equation}
each form a right moving ${SU(1|1)}$ subalgebra with the common central charge
$ \tilde L_0 - \tilde J^3$. We use the  $\tilde{}$
 to denote the generator corresponding to  the right movers.
The generators in (\ref{subgroup-1}) and ( \ref{subgroup-2})  annihilate the  chiral primaries of the
symmetric product CFT.  Let us denote the chiral primary with $L_0 = J^3, \tilde L_0 = \tilde J^3$ with
left and right $J^3$ charge $( \frac{J-1}{2}, \frac{J-1}{2})$ as
\begin{equation}
 |0\rangle_J \otimes |0\rangle_J.
\end{equation}
These are the ground states of the $\ZZZ_J$ twisted sector.
The magnons which are of interest in this paper are the following excitations above this
chiral primary.
\begin{equation}
\label{def-magnon}
 |\phi_{p_1} \phi_{p_2} \cdots \phi_{p_j} \rangle _J \otimes
|0\rangle_J = J_{p_1}^-J_{p_2}^- \cdots J_{p_j}^- |0\rangle_J \otimes |0\rangle_J
\end{equation}
with $J$ large and $J_p^-$ is given by
\begin{equation}
 J_p^- = \sum_{k=1}^J e^{ipk} J_{(k)}^-,
\end{equation}
and $J_{(k)}^-$ is the lowering operator of the left moving $SU(2)$  $R$-current of the
$k$-th copy
of the torus involved in the $\ZZZ_J$ twisted sector.
To satisfy orbifold group invariance condition we need to impose the condition
\begin{equation}
\label{physical}
 \sum_i p_i =0.
\end{equation}

At the free orbifold point of the symmetric product,
the magnon states in (\ref{def-magnon}) are non-chiral in the
left moving sector while it is still chiral in the right moving sector.
On perturbing the symmetric product CFT by the
 marginal operator which is a singlet of both the $SU(2)_R$ and $SU(2)_I$
 belonging to the $\ZZZ_2$ twisted sector,
  the magnons in ( \ref{def-magnon}) pick up anomalous dimensions.
The anomalous dimensions of these operators  was evaluated to the leading order in
conformal   perturbation theory  by \cite{Gava:2002xb} using methods developed
in \cite{Arutyunov:1997gi,Jevicki:1998bm,Lunin:2000yv,Lunin:2001pw}
\footnote{See \cite{Gomis:2002qi,Lunin:2002fw,Hikida:2002in} for related work. }.
  In \cite{David:2008yk} it was argued
that these magnons are BPS in a centrally extended
$SU(1|1) \times SU(1|1)$ algebra.
The dispersion relation of these magnons was obtained as a result of the BPS condition
in this extended algebra. To make this paper self contained,
 Appendix A. contains a brief review  of this result.
In this section it suffices to introduce this centrally extended algebra.
The generators of the  centrally extended $SU(1|1) \times SU(1|1)$
are composed of the set
\begin{equation}
 \{ G_{-1/2}^{++}, G_{1/2}^{--},  L_0 - J^3 \} \quad\hbox{and}\quad
\{ \tilde G_{-1/2}^{+-}, \tilde G_{1/2}^{-+}, \tilde L_0 - \tilde J^3 \}.
\end{equation}
To unclutter our notation we will define the following generators.
\begin{eqnarray}
\label{unclutter}
 G_{-1/2}^{++} \rightarrow Q_1, &\quad&
   \tilde G_{-1/2}^{+-} \rightarrow Q_2, \\ \nonumber
G_{1/2}^{--} \rightarrow S_1, &\quad& \tilde G_{1/2}^{-+} \rightarrow S_2, \\ \nonumber
L_0 - J^3 \rightarrow C_1, &\quad& \tilde L_0 - \tilde J^3 \rightarrow C_2.
\end{eqnarray}
Then the centrally extended algebra is given by
\begin{eqnarray}
\label{ext-algeb}
 \{Q_1, S_1\} = C_1, &\qquad& \{Q_2, S_2\} = C_2, \\ \nonumber
\{ Q_1, Q_2\} = C_3 - i C_4, &\qquad& \{ S_1, S_2\} = C_3 + i C_4, \\ \nonumber
\{Q_1, S_2 \} =0, &\qquad& \{ S_1, Q_2 \} =0.
\end{eqnarray}
where $C_3, C_4$ are the  two additional central charges.
Note that
\begin{equation}
 Q_a^\dagger = S_a, \qquad a \in \{1, 2\}
\end{equation}
This central extension of $SU(1|1)\times SU(1|1)$ can be viewed as a ${\cal N}=2$ Poincar\'{e} superalgebra
in $3$-dimensions with one central charge \cite{David:2008yk}.
The BPS condition for the extended algebra is given by
\begin{equation}
\label{bps-cond}
 \frac{1}{4} ( C_1 + C_2)^2 = \frac{1}{4} ( C_1 - C_2)^2 + C_3^2 + C_4^2 .
\end{equation}
Consider a single magnon state given by
\begin{equation}
|\phi_p\rangle_J\otimes |0\rangle_J = J_p^- |0\rangle_J \otimes |0\rangle_J.
\end{equation}
In \cite{David:2008yk} it was  argued  that
this magnon is a BPS state and carries the following values
of the central charges in the extended algebra  (\ref{ext-algeb})
\begin{eqnarray}
& & C_3 - iC_4 = \alpha( e^{-ip}  -1), \qquad
C_3 + i C_4 = \alpha^*(  e^{ip} -1), \\ \nonumber
&&C_1 -C_2 = ( L_0- J^3) - ( \tilde L - \tilde J^3) = 1,
\end{eqnarray}
where $\alpha$ is a function of the coupling of the strength of the marginal deformation of the
$\ZZZ_2$ twist operator.
Now from the  BPS condition in (\ref{bps-cond}) we obtain the following dispersion relation
\begin{eqnarray}
\label{dispersion-rel}
 C_1 + C_2 &=& ( L_0 + \tilde L_0) + ( J^3 + \tilde J^3), \\ \nonumber
&=& \Delta - J, \\ \nonumber
&=& \sqrt{ ( C_1 -C_2)^2 + 4C_1 C_2 }, \\ \nonumber
&=& \sqrt{ 1 + 16 g^2 \sin^2 ( \frac{p}{2}) }, \qquad g = |\alpha|.
\end{eqnarray}
$g$ will depend on the parameters of the D1-D5 system, namely the charges
$Q_1, Q_5$ and the six dimensional string coupling $g_6$.
In the next subsection we will mention on how it depends on these parameters.

For the purposes of this paper it will be useful to
 parametrize the dispersion relation using spectral parameters as follows: introduce
\begin{equation}
\label{xpm}
\frac{x^+}{x^-} = \exp(ip)
\end{equation}
subject to the constraint
\begin{equation}
\label{xpm-1}
 x^+ + \frac{1}{x^+} - x^- - \frac{1}{x^-} = \frac{i}{g}.
\end{equation}
We can use this constraint to evaluate the following
\begin{eqnarray}
\label{solxpm}
 x^- &=&\frac{i}{2g ( e^{ip} -1)} \left( 1 +   \sqrt{1 +  16 g^2 \sin^2 \frac{p}{2} } \right) , \\ \nonumber
c &=&  -i( x^+ - x^-) = \frac{1}{2 g}\left( 1+   \sqrt{1 + 16 g^2 \sin^2 \frac{p}{2} }  \right).
\end{eqnarray}
here we have taken  the positive branch in the square root.
We can therefore identify the central charges
\begin{eqnarray}
\label{central-ch}
& &  C_1 + C_2 = 2gc -1, \qquad
C_3 - i C_4 = \alpha ( \frac{x^-}{x^+} - 1 ) , \qquad \\ \nonumber
&& C_3 + i C_4 = \alpha^*  ( \frac{x^+}{x^-} -1),  \qquad
C_1 = gc, \qquad C_2 = gc -1.
\end{eqnarray}

There are other magnon like states which satisfy the BPS relation.
They carry the central charges
\begin{equation}
C_3 - i C_4 = \alpha( e^{ip} -1), \qquad C_3 + i C_4 = \alpha^*( e^{-ip} -1) ,
\qquad  C_1 - C_2 = Q.
\end{equation}
These can be thought of bound states of $Q$   elementary  magnons
and are called dyonic magnons.
Similar to the case of the elementary
giant magnons, their dispersion relation is given by
\begin{equation}
\label{dyon-disp}
C_1 + C_2 = \sqrt{ Q^2 + 16 g^2 \sin^2(\frac{p}{2} ) }.
\end{equation}
One can use spectral parameters to parametrize dispersion relation as
\begin{equation}
\frac{x^+}{x^-} = \exp(ip),
\end{equation}
with the constraint
\begin{equation}
\label{dyon-const}
x^+ + \frac{1}{x^+} - x^- - \frac{1}{x^-} = i \frac{Q}{g}.
\end{equation}
Then the dispersion relation is  given by
\begin{equation}
C_1 + C_2 = 2gc -Q,
\end{equation}
where $c$ is defined as $ c=   -i( x^+ - x^-)$.

\subsection{Magnons at strong coupling}

We have seen that  magnons are BPS states and they carry
large $J$ charge in the CFT, therefore we should expect to find them
as classical solutions to the string sigma model on $AdS_3\times S^3 \times T^4$.
Since the magnons have angular momentum along $S^3$ these solutions
are rotating along a direction in $S^3$.
Magnon solutions found for the case of $AdS_5\times S^5$ which have non-trivial
configurations along time and a $S^3$ within $S^5$ serve as magnon solutions
for the $AdS_3\times S^3$ case.  The description of magnons as semi-classical solutions makes
sense when the radius of curvature in comparison with string length
$R^2/\alpha' = \sqrt{g_6^2 Q_1Q_5} >>1$.
There are three interesting limits of these solutions.
\begin{enumerate}
\item{ \bf Plane wave limit} \\
This limit was discovered by \cite{Berenstein:2002jq} and is given by
\begin{equation}
 g\rightarrow \infty,\qquad  k = 2 gp \;\; \mbox{fixed}, \qquad Q\;\; \mbox{fixed}.
\end{equation}
The dispersion relation (\ref{dyon-disp}) then reduces to
\begin{equation}
\Delta - J = \sqrt{ Q^2 + k^2}.
\end{equation}
For a single magnon $Q=1$,
from ( \ref{solxpm}) we see that spectral parameters in this limit are given by
\begin{eqnarray}
x^+ \sim x^-& =& r + O(1/g), \\ \nonumber
&=& \frac{1}{k} \left( 1 + \sqrt{ 1+ k^2} \right) + O(1/g).
\end{eqnarray}
Thus the spectral parameters are real in this limit. From the above equation
one can re-write the momentum $k$ and the
frequency $\omega = \sqrt{1+ k^2}$  of the plane wave in terms of the
spectral parameter. This is given by
\begin{equation}
\label{defkom}
k(r) = \frac{2r}{ r^2 -1}, \qquad \omega(r) = \frac{r^2 +1}{r^2 -1}.
\end{equation}
These solutions are quanta associated with linearised fluctuations of the
world sheet fields around a string which orbits the equator of $S^3$
\cite{Berenstein:2002jq}
The fluctuations are of the form of plane waves and solve the linearised
equations of motion of the world sheet theory.
They have wave number $k(r)$ and frequency $\omega(r)$ given by
(\ref{defkom}).
States with $Q>1$ are bound states of plane waves.

 \item {\bf Giant magnon limit} \\
These solutions are obtained in the following limit
\begin{equation}
\label{giant-lim}
 g\rightarrow \infty, \qquad p  \;\;\mbox{fixed}, \qquad Q \;\; \mbox{fixed}.
\end{equation}
From (\ref{xpm-1}) and ( \ref{dyon-const})  we see that that  in this limit the
spectral parameters  have the property
\begin{equation}
 x^+ \sim \frac{1}{x^-} \sim \exp( ip/2) + O(1/g).
\end{equation}
Giant magnons solutions of the string sigma model for $AdS_3\times S^3\times T^4$ were studied in
\cite{David:2008yk} by using the giant magnon solution for the case of
$AdS_5\times S^5$
found by \cite{Hofman:2006xt}. They were shown to be BPS solutions in $AdS_3\times S^3\times T^4$
and their dispersion relation is given by
\begin{equation}
\label{lead-disp}
 \Delta- J =
\frac{R^2}{\pi\alpha'} \left| \sin\frac{p}{2} \right| +O( ( \frac{R^2}{\pi\alpha'} )^{0} ).
\end{equation}
From comparison with the exact dispersion relation in (\ref{dispersion-rel})
and the strong coupling dispersion relation (\ref{lead-disp})
 we see that $g$ is a function of the parameters  of the D1-D5 system
$g_6^2, Q_1, Q_5$ such that
\begin{equation}
\label{strong-rel}
 16 g^2 = \frac{R^2}{\pi\alpha'}
 = \frac{ g_6^2 Q_1 Q_5}{\pi^2} , \quad \mbox{for} \quad {g_6^2Q_1Q_5} >>1.
\end{equation}
Since these are classical solutions of the sigma model, corrections
to the dispersion relation will be
organized as inverse powers of $ \frac{R^2}{\pi\alpha'}$. We  perform a
one loop calculation and show that there is no
term of $O( ( \frac{R^2}{\pi\alpha'} )^{0} ) $
It is important to note that though the plane wave excitation and the
giant magnon classical solution looks different, they are representatives
of the same state in different regions of momentum space \cite{Maldacena:2006rv}.
\item
{\bf Dyonic giant magnon limit}\\
This limit is given by
\begin{equation}
 \label{dyon-lim}
 g\rightarrow \infty, \qquad Q\rightarrow\infty, \qquad
 \frac{Q}{g} = \mbox{fixed}, \quad p = \mbox{fixed}.
 \end{equation}
These solutions were found by
\cite{Chen:2006gea,Arutyunov:2006gs,Spradlin:2006wk,Minahan:2006bd}. They
have non-trivial field configurations in the $S^3$,  details of these solutions
are given in Appendix A.
Again from the constraint ( \ref{dyon-const}) obeyed by the spectral parameters
we find that
\begin{equation}
 x^+ = \bar x^- \sim O( g^{0}) .
\end{equation}
 The spectral parameters in this case depends on the parameter
$Q/g$ which can be tuned to any value, this fact will play an important
 role in our analysis.
The dispersion relation obeyed by the giant dyonic magnons is given by
\begin{equation}
\label{dyon-sdisp}
 \Delta - J = \sqrt{ Q^2 + \frac{R^2}{\alpha'} \sin^2 \left( \frac{p}{2} \right) }  +
O( ( \frac{R^2}{\pi\alpha'} )^{0} ).
\end{equation}
The dyonic giant magnon dispersion relation can be obtained from the
exact dispersion relation (\ref{dyon-disp}) by performing the
giant magnon limit. On comparing the result with the dyonic
giant magnon dispersion relation in (\ref{dyon-sdisp}) we obtain the
identification (\ref{strong-rel}). We will show that the correction to the dispersion relation (\ref{dyon-sdisp}) at one loop in the sigma model vanishes.
Both the giant magnon solution as well as the plane wave excitations
can be obtained as a further limit of the dyonic giant magnon.
$Q\rightarrow 0$ limit reproduces the giant magnon, and taking the limit
$x^+\sim x^-= r$ on the spectral parameters of the dyonic giant magnon
reduces it to the plane wave excitation  \cite{Chen:2006gea}.
\end{enumerate}

Studying the dispersion relation of magnons at strong coupling
helped us to make the identification (\ref{strong-rel}) between the
parameter $g$ and the parameters of the D1-D5 system. A similar comparison
can be done at weak coupling.
In \cite{Gava:2002xb}  magnons were  studied in the limit of small momentum $p$ and in first
order in the $\ZZZ_2$ blow up mode.  In this limit the dispersion relation of a single
magnon is shown to be
\begin{equation}
 \Delta -J = 1 + \frac{1}{2\pi^2} g_{6}^2 ( Q_1Q_5) \frac{p^2}{4},
\end{equation}
where $g_6$ is the coupling of the marginal $\ZZZ_2$ blow up mode in the
symmetric product.  Thus comparing this to the exact dispersion relation
(\ref{dispersion-rel})
we see that again we have
\begin{equation}
 16 g^2 = \frac{ g_6^2 Q_1 Q_5}{\pi^2}, \quad  \mbox{for} \quad g_6^2Q_1Q_ 5 <<1.
\end{equation}
A simple conjecture for the full dependence of the coupling $g$ on the parameters of the system
is then
\begin{equation}
\label{exact-rel}
16 g^2 = \frac{ g_6^2 Q_1 Q_5}{\pi^2}.
\end{equation}
We will test the  relation ( \ref{exact-rel}) to one loop in the strong coupling expansion.
Since perturbation
theory of the string sigma model is controlled by the parameter
$
\frac{R^2}{\pi \alpha'}
$
it is  clear that  possible corrections to the
relation ( \ref{exact-rel})
are of the form
\begin{equation}
4g = ( \frac{R^2}{\pi\alpha'} ) + g_0
+ g_1 (\frac{R^2}{\pi \alpha'})^{-1} + \cdots.
\end{equation}
Substituting this in the exact form of the dispersion relation  we obtain
\begin{equation}
\label{strong-exp}
E-J =\frac{R^2}{\pi\alpha'} \left| \sin\frac{p}{2} \right| + g_0 \left| \sin\frac{p}{2} \right|
+ f_1(p) (\frac{R^2}{\pi \alpha'})^{-1} + \cdots.
\end{equation}
We will evaluate the one loop correction to the dispersion relation at strong coupling
and show that $g_0=0$, which is consistent with the proposed relation (\ref{exact-rel}).

\section{The  $SU(1|1)$ invariant S-matrix for magnons}

In this section we show that demanding  $SU(1|1)$ symmetry,
the S-matrix for the scattering of magnons in the D1-D5 system
is determined up to a phase.
Recall that the symmetries preserved by the chiral primaries of the
D1-D5 system are two copies of $SU(1|1)\times SU(1|1)$ with common
central charges.
We demand that the S-matrix is symmetric under one of the $SU(1|1)$
in each copy of $SU(1|1)\times SU(1|1)$  using the
conventional co-product between two states.
Therefore let us focus on the $SU(1|1)$ generated by
$Q_1, S_1$ with central charge $C_1$.
The algebra is given by
\begin{equation}
 \{Q_1, S_1\} = C_1.
\end{equation}
To write down the $SU(1|1)$ invariant S-matrix we follow \cite{Beisert:2005wm}. Introduce the generator
$B$ which has the following commutation relations with the $SU(1|1)$ generators
\begin{equation}
 [B, Q_1] = -2Q_1, \qquad [B, S_1] = 2S_1.
\end{equation}
Basically $B$ is a $U(1)$ outer automorphism under which $Q_1$ and $S_1$
have charges $-2$ and $+2$ respectively.
This algebra has a quadratic Casimir given by
\begin{equation}
 {\cal J} = 2[Q_1, S_1] + \{B, C_1\}.
\end{equation}
Let us write down the action of these generators of one of the $SU(1|1)$ super algebra on
a single magnon which is  in the fundamental short multiplet of $SU(1|1)\times SU(1|1)$.
\begin{eqnarray}
\label{act-state}
 B|\phi_p\rangle \otimes |0\rangle = (\beta+1) | \phi_p\rangle \otimes |0\rangle, &\qquad&
B|\psi_p \rangle \otimes |0\rangle = ( \beta-1) |\psi_p\rangle\otimes |0\rangle, \\ \nonumber
Q_1 |\phi_p\rangle \otimes |0\rangle  = a|\psi_p\rangle\otimes |0\rangle, &\qquad&
Q_1 |\psi_p\rangle\otimes |0\rangle =0, \\ \nonumber
S_1  |\phi_p\rangle \otimes |0\rangle =0, &\qquad&
S_1 |\psi_p\rangle\otimes |0\rangle = b |\phi_p \otimes |0\rangle, \\ \nonumber
C_1 |\phi_p\rangle \otimes |0\rangle = ab  |\phi_p\rangle \otimes |0\rangle, &\quad&
C_1 |\psi_p\rangle\otimes |0\rangle = ab |\psi_p\rangle\otimes |0\rangle, \\ \nonumber
{\cal J}^2 |\phi_p\rangle \otimes |0\rangle  = 2 \beta ab |\phi_p\rangle \otimes |0\rangle ,
&\quad&
{\cal J}|\psi_p\rangle\otimes |0\rangle=
2\beta ab |\psi_p\rangle\otimes |0\rangle.
\end{eqnarray}
From the definition of the central charge $C_1$ in ( \ref{central-ch}) we obtain
\begin{equation}
ab = g c.
\end{equation}
Following the notation of \cite{Beisert:2005wm} we denote the above representation as
$({\bf 1}| {\bf 1})_{c, \beta}$.
The tensor product of two fundamental representations of $SU(1|1)$
decomposes into a direct sum of two fundamental representations.
We write this as
\begin{equation}
 ({\bf 1}| {\bf 1})_{c_1, \beta_1} \otimes ({\bf 1},|{\bf 1})_{c_2, \beta_2}
=
({\bf 1}| {\bf 1})_{(c_1+c_2), \beta_1 +\beta_2 +1} \oplus
({\bf 1}| {\bf 1})_{(c_1+c_2), \beta_1 +\beta_2 -1}.
\end{equation}
Since there are only two multiplets  occurring in the decomposition,  one only needs the
identity and the quadratic Casimir on the tensor product to write down any operator
which is invariant under the sum of the generators.
The quadratic Casimir on the tensor product is given by
\begin{equation}
\label{def-cas}
 {\cal J}^{(12)} =
2 g\beta_1 c_1 + 2 g\beta_2 c_2 +  2 B^{(1)} C_1^{(2)} + 2 C^{(1)}_1 B^{(2)} +
4 Q_1^{(1)} S_1^{(2)} - 4 S_1^{(1)} Q_1^{(2)}.
\end{equation}
We also need  the square of the quadratic Casimir which is given by
\begin{equation}
\label{def-casq}
 ( {\cal J}^{(12)} )^2 =  4(\beta_1 + \beta_2)( gc_1 + gc_2 ) {\cal J}^{(12)}
- 4 ( \beta_1 + \beta_2 + 1) ( \beta_1 + \beta_2 -1) ( g c_1 + gc_2)^2 {\cal I}^{(12)} .
\end{equation}
The $SU(1|1)$ S-matrix is written as the product of the graded permutation operator and the
R-matrix.
\begin{equation}
\label{def-smat}
 {\cal S} _{12} = {\cal P}_{12} { \cal R}_{12}( \alpha_1, \alpha_2) ,
\end{equation}
where $\alpha_1, \alpha_2$ are spectral parameters corresponding to the
two states.
Since the $R$-matrix is also an invariant under the sum of the generators
corresponding to the two sates, we can write it as
\begin{eqnarray}
\label{def-rmat}
 {\cal R}_{12}(\alpha_1, \alpha_2) = R_{12, 1}( \alpha_1, \alpha_2) {\cal I}^{(12)} + R_{12,2}(\alpha_1, \alpha_2)
{\cal J}^{(12)} .
\end{eqnarray}
We now demand that the $R$-matrix satisfies the unitarity constraint and the
Yang-Baxter relations given by the equations
 \begin{equation}
 {\cal R}_{12} {\cal R}_{21} = {\cal I}_{12}, \qquad
{\cal R}_{12} {\cal R}_{13} {\cal R}_{23} = {\cal R}_{23} {\cal R}_{13} {\cal R}_{12}.
\end{equation}
These conditions determine the scalars $R_{12,1}, R_{12, 2}$ to be of the form
\begin{eqnarray}
\label{def-rmat2}
R_{12,1}(a_1, a_2) &=& \frac{\alpha_2-\alpha_1 - \frac{i}{2}(\beta_1+\beta_2)(gc_1+gc_2) }
{\alpha_2 -\alpha_1 - \frac{i}{2} (gc_1 + gc_2)}
R_{12, 0} (\alpha_1, \alpha_2), \\ \nonumber
R_{12,2} &=& \frac{\frac{i}{4} }{ \alpha_2 -\alpha_1 - \frac{i}{2} (gc_1 +gc_2) } R_{12, 0}(\alpha_1, \alpha_2).
\end{eqnarray}
From the unitarity condition it is easy to see that
the undermined scalar $R_{12, 0}$ satisfies the condition
\begin{equation}
\label{phaser}
R_{12,0}(\alpha_1, \alpha_2) R_{12,0}(\alpha_2, \alpha_1) =1.
\end{equation}
Since the S-matrix is written in terms of the quadratic casimir of  the first
$SU(1|1)$, it is  has the following invariance
\begin{eqnarray}
\label{triv-cop}
 [Q_1^{(1)}\otimes 1 + (-1)^F\otimes Q_1^{(2)} , {\cal S}_{12} ] = 0, \\ \nonumber
[S_1^{(1)} \otimes 1 + (-1)^F \otimes S_1^{(2)}, {\cal S}_{12} ] = 0.
\end{eqnarray}
Thus this $SU(1|1)$ invariance is realized under the conventional co-product.
The symmetry corresponding to the
other $SU(1|1)$ in each copy is possibly realized using a non-trivial co-product.
In appendix A we write down a non-trivial co-product under which the
S-matrix written down in (\ref{def-smat}),  ( \ref{def-rmat}) (\ref{def-rmat2})
 is invariant with respect to the second $SU(1|1)$
in each copy.

Therefore we have shown that the requirement of invariance
 under one of the $SU(1|1)$  determines the S-matrix  up to a phase.
Let us write down the action of the S-matrix explicitly on the
two particle states. We first identify the  parameters $\alpha$ in terms of the
spectral parameters $x^\pm$ as
\begin{equation}
 \alpha = \frac{g }{2} ( x^+ + x^-) , \qquad \mbox{and} \qquad c = -i( x^+ - x^-).
\end{equation}
where the spectral parameters $x^+, x^-$ are related to the momentum $p_1$
of the first magnon
by  ( \ref{xpm}) and ( \ref{xpm-1}).
To write down the explicit action of the S-matrix we first define the its action on  two particle states  as
\begin{equation}
\label{smat-def2}
 {\cal S}_{12} |i \rangle_{(1)}  \otimes |j\rangle_{(2)}   =
S(p_2, p_1)_{ij}^{kl}|k\rangle_{(2)} \otimes |l\rangle_{(1)}.
\end{equation}
Now  given (\ref{def-smat}), ( \ref{def-rmat}) and ( \ref{def-rmat2}), we see that
explicitly the action of the S-matrix on the two magnon states is given by
\begin{eqnarray}
\label{explict-act}
 {\cal S}_{12}| \phi_{p_1} \phi_{p_2} \rangle \otimes |0\rangle
&=& \frac{y^+ - x^- }{ y^- - x^+} | \phi_{p_2} \phi_{p_1} \rangle \otimes |0\rangle, \\ \nonumber
{\cal S}_{12} |\phi_{p_1}  \psi_{p_2}\rangle \otimes |0\rangle &=&
\frac{ y^+ - x^+}{y^- - x^+} |\psi_{p_2} \phi_{p_1} \rangle \otimes |0 \rangle
+ \frac{y^+ - y^-}{y^- - x^+} \frac{ a^{(1)} }{a^{(2)}} | \phi_{p_2} \psi_{p_1}\rangle
\otimes |0\rangle, \\ \nonumber
{\cal S}_{12} |\psi_{p_1}\phi_{p_2}\rangle\otimes |0\rangle
&=& \frac{ y^- - x^-}{y^- - x^+} |\phi_{p_2} \psi_{p_1} \rangle\otimes
|0\rangle + \frac{x^+ - x^-}{ y^- - x^+} \frac{a^{(2)}}{ a^{(1)}}
| \psi_{p_2} \phi_{p_1}\rangle \otimes |0\rangle, \\ \nonumber
{\cal S}_{12} |\psi_{p_1}\psi_{p_2} \rangle \otimes |0\rangle
&=& -  | \psi_{p_2} \psi_{p_1}\rangle
\otimes |0\rangle.
\end{eqnarray}
Here $y^\pm$ refer to the spectral parameter of the second magnon and we have
suppressed the overall phase factor for convenience of notation.  $a^{(1)}, a^{(2)}$ refer
to the parameter $a$ in (\ref{act-state}) for the two states.
Note that since there are  two copies of the extended $SU(1|1) \times  SU(1|1)$ which share the same
central charge,  the full S-matrix is a tensor product of the both given by
${\cal S}_{12} \otimes {\cal S}_{12}$.  In this paper we will be interested in
magnons only in the $SU(2)$ sector.  We  can read out the scattering amplitude of these magnons
from the action of the  S-matrix   on the bosonic state $|\phi_1\phi_2\rangle$.
From the first line of (\ref{explict-act}) we see that
 the amplitude for scattering of two magnons in this sector is given by
\begin{equation}
\label{su2-smat}
 S(x^\pm, y^\pm)_{SU(2)} = S_0(x^\pm, y^\pm) \left( \frac{ x^+ - y^-}{x^- - y^+} \right)^2,
\end{equation}
where $S_0(x^\pm, y^\pm)$ is the undetermined phase factor.
$x^\pm, y^\pm$ implicitly
depend on momenta $p_1, p_2$ through the equations  (\ref{xpm}) and (\ref{xpm-1}).
We have also squared the amplitude due to the
existence of two copies of the extended  $SU(1|1) \times  SU(1|1)$.
The S-matrix found by imposing $SU(1|1) \times  SU(1|1)$ symmetry dictates the scattering 
of magnons in $AdS_3\times S^3$ with no excitations on $T^4$. 
Because the $T^4$ worldsheet theory is decoupled, the only possible coupling between these 
excitations and that of the $T^4$ is the virasoro constraint. Thus the excitations 
on $T^4$ can be consistently set to zero and work with a consistent subsector in which 
one examines the excitations only on $AdS_3\times S^3$. 

\subsection{Leading contribution to the phase factor}

The phase factor $S_0(x^\pm, y^\pm)$
in general cannot be determined by symmetry considerations alone. In this paper we will evaluate the
phase factor to the leading and the first sub-leading terms in the semi-classical expansion.
 For this it is convenient to parametrize
the phase factor $S_0(x^\pm, y^\pm)$ as follows:
\begin{eqnarray}
\label{phase}
 S_0(x^\pm, y^\pm)
 &=& \sigma_{{\rm BDS}} \times \sigma^2(x^\pm, y^\pm ), \\ \nonumber
 &=& \frac{ x^- - y^+}{ x^+ - y^-} \frac{ 1 - \frac{1}{x^+ y^-} }{ 1 - \frac{1}{x^- y^+}}
\times \sigma^2(x^\pm, y^\pm ),
\end{eqnarray}
where
\begin{equation}
\label{bds}
\sigma_{\rm BDS} =
 \frac{ x^- - y^+}{ x^+ - y^-} \frac{ 1 - \frac{1}{x^+ y^-} }{ 1 - \frac{1}{x^- y^+}}.
\end{equation}
This parametrization of the undetermined phase is used so that  it  is
easy to see the difference from the scattering amplitude of
two magnons in the $SU(2)$ sector of ${\cal N}=4$ Yang-Mills.
In that theory, the scattering amplitude of the $SU(2)$ subsector is of the same
form as in ( \ref{su2-smat}) and (\ref{phase}) with the same pre-factor $\sigma_{{\rm BDS}}$
\cite{Beisert:2004hm}.
the possible difference between the two theories is
 therefore parametrized  by differences in  $\sigma$.
Let us write $\sigma$ as
\begin{equation}
\label{def-sig}
 \sigma (x^\pm, y^\pm) = \exp( i\theta (x^\pm, y^\pm) ).
\end{equation}
Unitarity demands that $\theta(x,y)$ is anti-symmetric in its arguments.
We further write $\theta$ in terms of the strong coupling expansion as
\begin{equation}
\label{phase-exp}
 \theta(x^\pm, y^\pm) =  g \left( \theta_0(x^\pm, y^\pm) +
 \frac{1}{ g } \theta_1(x^\pm, y^\pm) +
\frac{1}{ g^2} \theta_2(x^\pm, y^\pm) + \cdots  \right).
\end{equation}
This form of  the expansion is easily motivated by examining the sigma model
expansion which is organized in terms of inverse powers of
 $\frac{R^2}{\pi \alpha'}$,
 with the leading term proportional to $\frac{R^2}{\pi\alpha'}$ and the fact that at strong coupling
$4g =\frac{R^2}{\pi \alpha'}$.

The leading semi-classical contribution $\theta_0(x^\pm, y^\pm)$ can be evaluated
using the relation between phase shift and time delay for the scattering of
 two magnons given by \cite{Jackiw:1975im}
\begin{equation}
\frac{\partial \theta_0(p_1, p_2) }{\partial{E_{p_1}} }= \Delta T_{12},
\end{equation}
where $E_{p_1}$ is the energy of the first magnon.
This leading term in the S-matrix for the scattering of two magnons has been
evaluated  in \cite{Hofman:2006xt}.  Though this was evaluated for
giant magnons in the $AdS_5\times S^5$ geometry, the answer just depends on the classical solution
of the magnons. In fact the crucial ingredient in the calculation was just the
map of classical string theory on $R\times S^2$ to the Sine-Gordan model.
 In the strong coupling limit, the magnon solution of the D1-D5 system is identical
to that of ${\cal N}=4$ Yang-Mills.  It is just a solution of
classical string theory on $R\times S^2$.
 Therefore the leading term
in the strong coupling expansion for  the S-matrix  for scattering of two
giant magnons with
momentum $p_1$ and $p_2$ is the same as that evaluated in \cite{Hofman:2006xt}.
This is given by
\begin{eqnarray}
\label{time-delay}
& &S(p_1, p_2) = \exp( i\delta), \\ \nonumber
&& \hbox{where} \;\; \delta =
- \frac{\sqrt{ g_6^2 Q_1Q_5} }{\pi} \left(  \cos \frac{p_2}{2} - \cos\frac{p_1}{2} \right)
\log\left( \frac{\sin^2 \frac{p_1-p_2}{4} }{ \sin^2 \frac{p_1+p_2}{4}} \right),
\end{eqnarray}
with sign$(\sin(\frac{p_1}{2} )>0$ and
sign$(\sin(\frac{p_2}{2} )>0$.
Thus   in the leading semi-classical
limit, the factor $\sigma(x, y)$ is identical to that evaluated for the case of
magnons in the ${\cal N}=4$ Yang-Mills. This is given by
\begin{equation}
\label{lead-phase1}
\theta_0(x^\pm, y^\pm) = k( x^+, y^+) - k(x^+, y^-) - k(x^-, y^+) + k( x^-, y^-),
\end{equation}
where
\begin{equation}
\label{lead-phase}
 k(x, y) = \left[ \left(  y + \frac{1}{y}\right)  - \left(x + \frac{1}{x} \right) \right] \log \left(
1 - \frac{1}{xy} \right) .
\end{equation}
This form of the leading contribution of the phase for the scattering of magnons in the $SU(2)$ sector for ${\cal N}=4$ Yang-Mills  in terms of the spectral parameters
was first proposed in \cite{Arutyunov:2004vx}.
To show that  this phase factor agrees with
that computed using the semi-classical solution of
giant magnons in $R\times S^2$,  all we have to do is to evaluate (\ref{su2-smat}) with
$S_0$ given by ( \ref{phase}) in terms of the momenta, rather than the spectral
parameters.
In the giant magnon limit (\ref{giant-lim})   we can
expand the equations (\ref{solxpm}) to  the leading order.
For sign$(\sin\frac{p_1}{2} )>0$,  sign$(\sin\frac{p_2}{2})>0$ we obtain
\begin{eqnarray}
 x^- = \exp( \frac{-ip_1}{2} )  + O(1/g) , \quad x^+ = \exp( \frac{ip_1}{2} ) + O(1/g) , \\ \nonumber
y^- =  \exp( \frac{-ip_2}{2} )  + O(1/g) , \quad y^+ = \exp( \frac{ip_2}{2} ) + O(1/g ).
\end{eqnarray}
Substituting this in the expression for the scattering amplitude  (\ref{su2-smat})
with $S_0$ given by ( \ref{phase})and retaining only
the leading order contribution from the phase (\ref{lead-phase}) we obtain
\begin{eqnarray}
 S(p_1, p_2)  &=& \exp ( i \delta ), \\ \nonumber
\delta  &=& -  \frac{\sqrt{g_6^2 Q_1Q_5}}{\pi}  \left(  \cos \frac{p_2}{2} - \cos\frac{p_1}{2} \right)
\log\left( \frac{\sin^2 \frac{p_1-p_2}{4} }{ \sin^2 \frac{p_1+p_2}{4} }\right).
\end{eqnarray}
Here we have substituted the relation $4g =R^2/\pi\alpha'$ which is valid in the
strong coupling limit.
Note that this agrees precisely with
(\ref{time-delay}),  the leading  amplitude obtained from using the
classical solution of the magnons in the $R\times S^2$ geometry.

\section{One-loop corrections}

Our goal in this  section is to determine the one-loop
correction to the phase factor $\theta_1(p_1, p_2)$
as well as the one-loop correction to the dispersion relation at strong coupling.
For the case of ${\cal N}=4$ Yang-Mills this phase was first evaluated in \cite{Hernandez:2006tk} 
in fact an all order expansion of this phase  consistent with crossing symmetry was 
proposed in \cite{Beisert:2006ib}. 
To evaluate $\theta_1(p_1, p_2) $  for our system
we follow the method  developed by \cite{Chen:2007vs}.
Let us  now summarize their approach:
Let $\delta(k;p)$ be the phase shift corresponding to the scattering of a plane wave
off the either a giant or a dyonic magnon. Where
$k$ is the momentum of the plane wave
of charge $Q$ and $p$ the momentum carried by the magnon.
Then the one-loop correction to the dispersion relation is given by
\cite{Chen:2007vs}
\begin{equation}
\label{1-ldisp}
 \Delta E(p)  = \frac{1}{2\pi} \sum_{I=1}^{N_F} (-1)^{F_I} \int _{-\infty}^\infty
dk \frac{\partial \delta_I (k;p)}{ \partial k} \sqrt{k^2 + Q^2}.
\end{equation}
Here $I$ labels the fluctuations of the magnons with Bose/Fermi statistics depending on the
sign of $(-1)^{F_I}$.
While the one loop correction to the scattering phase is given by
\begin{eqnarray}
\label{1-lphase}
\nonumber
\Delta\Theta (p_1, p_2)  = \frac{1}{4\pi} \sum_{I=1}^{N_F} ( -1)^{F_I} \int_{-\infty}^\infty
dk \left( \frac{\partial \delta_I(k; p_1)}{ \partial k} \delta_{I} ( k, p_2)
- \frac{\partial \delta_I(k; p_2)}{ \partial k} \delta_{I} ( k, p_1) \right). \\
\end{eqnarray}
The above formula is explicitly anti-symmetric in $p_1, p_2$, but the
second term is equal to the first one up to a total divergence \footnote{In \cite{Chen:2007vs}
the formula was written without the explicit anti-symmetrization.}.
Integrability of the world sheet sigma model is
an important ingredient which goes into the derivation  of the
 one loop phase shifts for the scattering  of two solitons.
\cite{Chen:2007vs}.

Both ( \ref{1-ldisp}) and ( \ref{1-lphase}) are functions of the momentum $p_1, p_2$
of the magnon, however in the expressions for the $S$-matrix
given in (\ref{su2-smat}) and  (\ref{phase}),
the expansion of the phase factor in (\ref{phase-exp})  are written in terms
of the spectral parameters $x^\pm, y^\pm$ of the magnon.
The spectral parameters for the elementary magnon
are related to the momentum by equations
(\ref{solxpm})  which involve the coupling $g$.
In general one might expect that on substitution of the spectral parameters of the
magnon in terms of their momenta there might be mixing between
what we have organized as  the BDS factor, $\theta_0$ and $\theta_1$.
If this occurs then the formula in (\ref{1-lphase}) would only give the
contribution of what occurs as the coefficient of $g^0$ term and would
not directly give $\theta_1$.
The way to avoid this difficulty is to use the expression (\ref{1-lphase})
for dyonic magnons \cite{Chen:2007vs}.
 Note that for dyonic magnons in the classical
limit (\ref{dyon-lim}),
the spectral parameters are of order one and due to the presence of the
additional parameter $Q$ (\ref{dyon-disp}).
 This ensures that in the expansion organized as
(\ref{phase-exp})  and there is no mixing of orders. There is a correction to
$\theta_0$ from the $\sigma_{\rm BDS}$ factor due to the fact that dyonic magnons are
bound states, however there is no change in the one loop
term $\theta_1$ \cite{Chen:2006gq,Chen:2007vs}.
Though this was found for $SU(2)$ giant magnons in
the case of $AdS_5$ the same arguments go through for the
the case of $SU(2)$ magnons in $AdS_3$ since this relies only on the
$\sigma_{\rm BDS}$  factor and $\theta_0$ which is identical for both.
In summary to read out the one loop correction $\theta_1$ we can
use the classical solutions of dyonic magnons and read out the plane wave  phase shifts
shifts $\delta(k, x^\pm)$ directly in terms of the spectral parameters of
the dyonic magnons and substitute them in (\ref{1-lphase}) to read out
$\theta_1$. Thus we have
\begin{equation}
\label{rel-phase}
2\theta_1( x^\pm, y^\pm)  = \Delta\Theta(p_1(x^\pm) , p_2 (y^\pm) ).
\end{equation}
Note that there is a factor of $2$ on the LHS of (\ref{rel-phase}) since this is
total one loop phase contribution in $\sigma^2(x^\pm, y^\pm)$.

Finally we mention that in  near horizon geometry   of the D1-D5 system
 there are excitations 
in $AdS_3\times S^3$ as well as along $T^4$.  
As we have mentioned earlier the magnons we consider 
 have polarizations only along $AdS_3\times S^3$ and we have set the excitations along $T^4$ to zero. 
This is certainly a consistent truncation as we will argue now. 
The bosonic and fermionic  co-ordinates for $T^4$ couple trivially to the
co-ordinates of $AdS_3\times S^3$.  The only possible coupling between these coordinates
arises due to the Virasoro constraints \footnote{We thank the referee 
for emphasizing this point.}.  These constraints state that
$T_{AdS_3\times S^3} + T_{T^4} =0$, where $T$ refers to the worldsheet stress tensor. 
If we do not consider any excitations along $T^4$,  then  the virasoro constraints  need to 
be only imposed on excitations along $AdS_3\times S^3$.  
Therefore we can consistently set the excitations along $T^4$ to zero. 
But, by doing this we do miss information about these states. 
Let us illustrate this by the following simple example. 
Consider a single magnon 
of momenutm $p$ but in the plane wave limit  in $AdS_3\times S^3$ and  excitations along 
$T^4$. The dispersion relation and the level matching constraints are given by \cite{Gava:2002xb}. 
\begin{eqnarray}
 \Delta - J &=& \sqrt{ 1 + 4 g^2 p^2} + g \frac{ L_0^{T^4} + \tilde L_0^{T^4}}{ J} , \\ \nonumber
p &=& \frac{ L_0^{T^4} - \tilde L_0^{T^4}}{J} 
\end{eqnarray}
Here $L_0$ is the zero mode of the stress tensor, note that these are the 
manifestations of the virasoro constraint in the light cone gauge.  The second equation is the condition 
that the total world sheet momentum vanishes. 
Thus we see that though we can consistently set the excitations along  $T^4$ to zero, we do miss 
the modifications to the dispersion relation due to the these excitations if they are turned on. 
This fact was observed in \cite{Babichenko:2009dk}. 
The formula in 
(\ref{1-ldisp}) and (\ref{1-lphase})  has been written with the restriction to the excitations
only along $AdS_3\times S^3$ which are all massive. 
 For the rest of the paper we ignore the $T^4$ directions and deal with only
$AdS^3\times S^3$. It will be certainly interesting to see if the semi-classical methods used in this 
paper can be extended to the case of excitations along $T^4$. In particular derive the above 
modifications to the dispersion relation and see if there are any modifications to the S-matrix
due to these excitations.

The rest of this section is organized as follows:
We first evaluate the phase shifts suffered by the plane waves
which are bosonic when they scatter off
the dyonic magnon solution by using the dressing method of
\cite{Spradlin:2006wk}. We then show that the phase shifts of the
fermionic plane wave excitations are $1/2$ of the bosonic  ones
using the finite gap method of \cite{Kazakov:2004qf}.
Finally we use (\ref{1-ldisp}) and ( \ref{1-lphase}) to obtain the one loop corrections
to the dispersion relation and the one loop correction
 to the  phase factor in the S-matrix.

\subsection{ Bosonic phase shifts: The dressing method}\label{dressmain}

As we have discussed earlier, the dyonic giant magnon solution reduces to the
plane wave excitation on taking the limit $x^+\sim x^- =r$.
The  corresponding plane wave excitation is a solution of the
linearised equation of motion with the following wave number and
frequency
\begin{equation}
\label{wavnu}
k(r) = \frac{2r}{r^2 -1}, \qquad \omega(r) =  \frac{r^2 +1}{r^2 -1}.
\end{equation}
This observation facilitates in extracting out the phase shift suffered
by a plane wave scattering off the dyonic giant magnon solution.
The strategy is a follows: The dressing method of \cite{Spradlin:2006wk} provides
a construction of a solution of $N$ dyonic giant magnons (DGMs)
which have non-trivial configurations in $S^3$. They are parametrized
by their spectral parameters $x_i^\pm$ and $ i = 1, \cdots N$.
We then use the dressing method again to obtain a solution
with $N+1$ DGMs and take the limit where this new DGM reduces to a
plane wave excitation. Thus we would have obtained a linearised fluctuation
around the $N$ DGM's. This provides an easy method to extract out its
spectrum as well as the phase shift suffered by the plane wave scattering
in the $N$ DGM background.

Since we are interested in a DGM in the $S^3$ of $AdS^3\times S^3$ we can
ignore the $AdS^3$ directions. The $S^3$ directions are parametrized as follows
\be \label{1.1}
\{(Z_1,Z_2):|Z_1|^2+|Z_2|^2=1\} \quad \leftrightarrow
\quad g=\left(
     \begin{array}{cc}
         Z_1 & -iZ_2 \\
       -i\bar{Z}_2 & \bar{Z}_1 \\
          \end{array}
     \right) \in SU(2),
\ee
where $g(z,\bar{z})$ is a $2\times 2$ matrix valued valued field satisfying the equations of motion
\be \label{eom}
\bar{\partial}\left(\partial g ~~ g^{-1}\right)+{\partial}\left(\bar{\partial} g ~~ g^{-1}\right)=0.
\ee
The dressing method provides a easy construction of the solution corresponding
to DGM's. Then by taking the plane wave limit on one of the DGM we
can obtain the bosonic phase shifts suffered by this plane wave as it
scatters off the rest of the DGMs. The details of this calculation is given in the
 appendix \ref{dressappendix}, we quote the final results.
From (\ref{1.17a}) we can read out the phase shifts for the
perturbation associated with the S$^3$ fluctuations around a $N$-DGM background.
This is given by
\ben \label{1.17}
\delta_{Z_1} \equiv -i\ln(\delta Z_1(\infty))+i\ln(\delta Z_1(-\infty))&=&0, \\ \nonumber
\delta_{Z_2} \equiv -i\ln(\delta Z_2(\infty))+i\ln(\delta Z_2(-\infty))&=&-\sum_{i=1}^N \left[  2i \ln\left(\frac{r- x_i^+}{r- x_i^-}\right)  -i
\ln\left(\frac{x^+_i}{x^-_i}\right)\right].
\een
where $r$ is the spectral
parameter associated with the plane wave which is related to the frequency $\omega$ and wave number $k$ by (\ref{wavnu}).
As is shown in (\ref{1.18a}) the phase shifts for the complex conjugate fields
are given by
\begin{eqnarray}
\label{1.18}
 \delta_{\bar Z_1}(1/r)  \equiv -i\ln(\delta \bar{Z_1}(\infty))+i\ln(\delta \bar{Z_1}(-\infty))&=&0, \\ \nonumber
\delta_{\bar Z_2}(1/r)   \equiv -i\ln(\delta \bar{Z_2}(\infty))+i\ln(\delta \bar{Z_2}(-\infty)) &=&\sum_{i=1}^N \left[  2i \ln\left(\frac{r- x_i^+}{r- x_i^-}\right)  -i
\ln\left(\frac{x^+_i}{x^-_i}\right)\right] .
\end{eqnarray}
A simple consistency check of these results for phase shifts is that
if the spectral parameter of any of the DGM reduces to a real number which means that the DGM
is in fact a plane wave excitation, then the phase shift must vanish. This is because
there is no scattering between two plane wave excitations. This property  can be  seen to be
true form the expressions in (\ref{1.17}) and (\ref{1.18}).
As we have discussed before, since the bosonic co-ordinates of $AdS_3$ couple
trivially to that of the $S^3$  we expect the plane wave fluctuations along
the $AdS_3$ directions to suffer no phase shifts. On the other hand the
fermionic co-ordinates of $S^3$ and that of $AdS_3$ couple with each other, therefore
we expect plane wave fluctuations along all the fermionic co-ordinates to suffer
phase shifts. We will show this explicitly in the next section using the finite gap
approach.

\subsection{Fermionic phase shifts: Finite gap method}

To obtain the phase shifts corresponding to the fermionic plane wave fluctuations
around the dyonic magnon and to prove that there are no phase shifts
for bosonic plane wave fluctuations in $AdS_3$ we use
 the description of
classical solutions with periodic boundary conditions developed in
\cite{Kazakov:2004qf} for the case of the sigma model on $AdS_3\times S^3$.
Our discussion closely follows that of $AdS_5\times S^5$ in \cite{Chen:2007vs}.
Classical string propagation on $AdS_3\times S^3$ with Ramond-Ramond flux can be described as a
non-linear sigma model on the supergroup  $SU'(1,1|2)$ \cite{Berkovits:1999im}.  The sigma model is given by
\begin{equation}
 S = - \frac{R^2}{2\pi \alpha'} \int d^2 z {\rm Tr}' [ \partial^\mu g^{-1} \partial_\mu g ] ,
\end{equation}
where $g$ takes values in the supergroup $SU'(1,1|2)$ and $\rm{Tr}'$ is the non-degenerate
bi-invariant metric.
The metric has the signature $(-1, 1, 1, 1, 1, 1)$\footnote{See \cite{Berkovits:1999im}
for details.}. We will not
require the detail structure of the sigma model but only the
properties of  group element $g$.
 An element of the  supergroup $SU'(1,1|2)$ satisfies the following
\begin{equation}
\label{param-1}
 g = \exp( i  x),
\end{equation}
where $x$ is a $4\times 4$ supermatrix given by
\begin{equation}
\label{param-2}
x = \left(\begin{array}{cc}
           a & b\\ c &d
          \end{array}
\right),
\end{equation}
with $a$ and $d$  being bosonic Hermitian $2\times 2$ matrices and $b$ and $c$ fermionic $2\times 2$ matrices such that $b = c^\dagger$.
They also satisfy
\begin{equation}
\label{parem-3}
{\rm{Tr}} \; a = {\rm{Tr}}\; b =0.
\end{equation}
The bosonic part of the supergroup $SU'(1, 1|2)$ is given by $SU(1,1) \times SU(2)$.
Classical solutions of the sigma model are described in terms of the monodromy
matrix constructed using the one form $j$ given by
\begin{equation}
 j = - dg g^{-1} .
\end{equation}
The equations of motion of the sigma model is given by
\begin{eqnarray}
\label{eom-1}
\partial_+ j_- + \partial_-j_+ =0,
\end{eqnarray}
where the $\pm$ subscripts refer to the light cone directions of the world sheet.
The one form $j$ also satisfies the identity
\begin{equation}
\label{flat-1}
\partial_+ j_- - \partial_-j_+ + [j_+, j_-] =0.
\end{equation}
It can then be shown  using (\ref{eom-1}) and ( \ref{flat-1}) the connection
\begin{equation}
J_{\pm}(x)  = \frac{ j_{\pm}( \sigma, \tau)}{ 1\mp x},
\end{equation}
with $x\in C$,  is a family of  flat connections.
That is it obeys the zero curvature condition
\begin{equation}
\partial_+ J_- - \partial_- J_+ + [J_+, J_-] = 0.
\end{equation}
Using this flat connection we can construct the monodromy matrix
\begin{equation}
\Omega(x) = P \exp\left[ \int_0^{2\pi}d\sigma \frac{1}{2} \left( \frac{j_+}{1-x} - \frac{j_-}{1+x}  \right) \right],
\end{equation}
where $P$ refers to path ordering.
Note that here $x$ is the spectral parameter which characterizes the flat connection.
Since the monodromy matrix takes values in the supergroup $SU'(1, 1|2)$
its eigenvalues are of the form
\begin{equation}
 \{ e^{i \hat p_1},  e^{i \hat p_2} | e^{i\tilde p_1}, e^{i \tilde p_2} \},
\end{equation}
with
\begin{equation}
\label{eigen-const}
\hat p_1 = - \hat p_2, \qquad \tilde p_1 = - \tilde p_2.
\end{equation}
The above condition arises from the definitions in  (\ref{param-1}), ( \ref{param-2})
and  the traceless property (\ref{parem-3}).
Classical strings propagating on the sigma model are classified by the
analytical properties of the eigenvalues of the monodromy matrix.
The $p$'s are the quasi-momentum, they are meromorphic functions of
the spectral parameter $x$. We will label them $p_{i}$ with $i=1, 2, 3, 4$ and
$p_1 = \hat p_1, p_2 =\hat p_2, p_3 = \tilde p_1, p_4 =\tilde p_2$. They have
the following properties \cite{Kazakov:2004qf}:
\begin{enumerate}
\item
$p(x)$ has poles with equal residue $-\frac{l}{2}$ at points
$x = \pm 1$ where $l$ is the length of the string.
\item $p(x)$ can have branch cuts in the complex plane, its discontinuity across
each cut is fixed by the equation
\begin{equation}
\label{fun-eq}
 p_i(x+i\epsilon) + p_j(x-i\epsilon) = 2\pi n_{ij},
\end{equation}
where $n_{ij} \in \ZZZ$.
\item
Using properties (1) and (2) we can write the quasi-momentum as
\begin{equation}
\label{def-quasi}
 p_i(x) = G_i(x) - \frac{l}{2} \left( \frac{1}{x-1} + \frac{1}{x+1} \right),
\end{equation}
where $G(x)$  is called the resolvant.
Substituting this form for the resolvant in (\ref{fun-eq}) we obtain the
fundamental equation for the resolvant given by
\begin{equation}
\label{fun-eq1}
 G_i(x+ i\epsilon) + G_j(x-i\epsilon)
= 2\pi n_{ij}  + l  \left( \frac{1}{ x-1} + \frac{1}{x+1} \right) .
\end{equation}
\item
The various different classical solutions satisfying the Virasoro conditions
are given by different solutions for the resolvant.
\item
Plane wave excitations about the classical solutions are described by
introducing a pole at a position $r$ with unit residue  on the real line.
The position of the pole is constrained by the equation
\begin{equation}
 G_i(r+i \epsilon) + G_j(r - i \epsilon)    = 2\pi n_{ij} +l
\left( \frac{1}{r-1} + \frac{1}{r+1} \right),
\end{equation}
where $G_i$ is the resolvant of the corresponding classical solution.
\item
 From (\ref{defkom}) we  can now identify  $k(r) = \frac{2r}{r^2-1}$
to be the wave number of the plane wave excitation and write the above equation as
\begin{equation}
\label{mom-q}
 G_i(r+i \epsilon) + G_j(r - i \epsilon)  - k(r) l  = 2\pi n_{ij} ,
 \qquad n_{ij} \in \ZZZ
 \end{equation}
This equation is the quantization condition for the plane waves and
it determines the phase shifts  suffered by the plane waves on scattering off the
soliton described by the resolvant $G_i$. The phase shifts corresponding to
plane wave excitation with $(ij)$  polarization is
 given by $ -(G_i(r+i \epsilon) + G_j(r - i \epsilon)) $
\end{enumerate}

We can now apply these results to the case of the dyonic giant magnon
and determine the  relation of the fermionic plane wave to the bosonic ones.
Let us first embed the $SU(2)$ dyonic magnon solution in $AdS_3\times S^3$.
As we have seen in the previous section,
this is a purely bosonic configuration of fields for which only the
bosons corresponding to the $SU(2)$ are turned on. Thus the resolvant
corresponding to the $\tilde p_1$ and $\tilde p_2$ quasi momentum
is non-zero. Using (\ref{eigen-const}) we have
\begin{equation}
 \tilde p_1(x) = -\tilde p_2(x) =   G(x)- l \frac{x}{x^2 -1}.
\end{equation}
Since there is no non-trivial configurations of fields turned on in the
other directions, the corresponding resolvants  vanish and the
quasi momenta have only the poles at $x=\pm 1$. Again using
(\ref{eigen-const}) we have
\begin{equation}
\label{adsmom}
 \hat p_1(x) = - \hat p_2(x) = - l \frac{x}{x^2 -1}.
\end{equation}

\noindent
We now can determine the phase shifts for various polarizations of the plane
wave excitations about the dyonic giant magnon using the relation (\ref{mom-q}).
\begin{enumerate}
\item
For fluctuations along the co-ordinate $Z_2$ in $S^3$, the plane wave
momenta is quantized as
\begin{equation}
\tilde p_1(r) -\tilde p_2(r) = 2 G(r) - k(r) l = 2\pi n_{\tilde 1\tilde 2},
\end{equation}
Thus the phase shift along the co-ordinate $Z_2$ is given by $- 2 G(r)$ where
$G(r)$ is the resolvant of the dyonic giant magnon solution.
\begin{equation}
\label{p1}
\delta_{Z_2} (r) = - 2 G(r).
\end{equation}
\item
The phase shift for plane wave fluctuations along
 the conjugate co-ordinate $\bar Z_2$ using this method is
given by
\begin{equation}
\tilde p_2(r) -\tilde p_1(r) = -2 G(r) + k(r) l = 2\pi n_{\tilde2\tilde1} ,
\end{equation}
To read out the phase shift one need to cast it same form as in
(\ref{mom-q}). We do this by using the relation
$k(r) = - k(1/r)$, then the above equation becomes
\begin{equation}
\tilde p_2(1/r) -\tilde p_1(1/r) = -2 G(1/r) - k(r) l = 2\pi n_{\tilde 2\tilde 1},
\end{equation}
Thus the phase shifts along the $\bar Z_2$ direction is given by
$2 G(1/r)$.
\begin{equation}
\label{p2}
\delta_{\bar Z_2}(r) = 2 G(1/r).
\end{equation}
\item
Plane wave phase shifts along the  complex conjugate pairs of
$AdS_3$ directions vanish, and that  can be seen from the
quantization condition
\begin{equation}
\hat p_1(r) - \hat p_2 (r) = k(r)l = 2\pi n_{\hat 1\hat 2}.
\end{equation}
Reading out the phase shift we have
\begin{equation}
\label{p3}
\delta_{Y_2}(r) = \delta_{\bar Y_2}(r)  =0.
\end{equation}
\item
There are 2 complex fermions
say $\theta$ and $\eta$ which are partners of the co-ordinates $Z_2$ in
$S^3$ and $Y_2$ in $AdS_3$. Phase shifts along these directions are given by the
quantization conditions
\begin{equation}
\tilde p_1 - \hat p_2 = G(r) - k(r) l = 2\pi n_{\tilde 1 \hat 2}, \qquad
\tilde p_2 - \hat p_1 = G(r) - k(r) l = 2\pi  n_{\tilde 2\hat 1}.
\end{equation}
Thus the phase shifts along the fermionic directions are given by
\begin{equation}
\label{p4}
\delta_{\theta} (r) =  \delta_{\eta} (r) = - G(r).
\end{equation}
\item
Going through a similar argument for the case of the bosons
we obtain the plane wave phase shifts along the complex conjugate
fermionic directions to be
\begin{equation}
\label{p5}
\delta_{\bar\theta} (r) =  \delta_{\bar \eta} (r) =  G(1/r).
\end{equation}
\end{enumerate}

We now have the phase shifts for plane wave fluctuations along all the $4$ transverse directions of
$AdS_3\times S^3$ in terms of the resolvant $G(r)$ of the dyonic magnon
solution.
 From the explicit calculation of the
bosonic phase shifts along the $Z_2$ and $\bar Z_2$  directions in (\ref{1.17}) and (\ref{1.18})
 we can identify $G(r)$ to be
\begin{equation}
G(r) = G(r, x^\pm) = - \left[ \frac{1}{i} \ln\left( \frac{r- x^+}{r -x^-} \right)  -
\frac{1}{2i} \ln\left(  \frac{x^+}{x^-}\right) \right].
\end{equation}
Here we have taken the value of the phase shift suffered by the plane wave on scattering off
a single giant magnon.
To summarize we have the following results for the plane phase wave shifts
along various polarizations
\begin{eqnarray}
\label{end-phase}
 S^3 &: &\delta_{Z_2} = - 2G(r, x^\pm), \quad \delta_{\bar Z_2} = 2 G(1/r, x^\pm), \\ \nonumber
AdS_3& :& \delta_{Y_2} = \delta_{\bar Y_2} =0, \\ \nonumber
\hbox{fermionic}&:& \delta_{\theta} = \delta_{\eta} = - G(r, x^\pm), \\ \nonumber
&:& \delta_{\bar \theta} = \delta_{\bar \eta} =  G(1/r, x^\pm).
\end{eqnarray}

\subsection{ One loop energy shift}

To evaluate the  one loop energy shift we
 can  substitute the phase shifts given  in (\ref{end-phase})
 into the expression for  the one loop correction for the
energy  of a magnon with momentum $p$ given in (\ref{1-ldisp}).
We obtain the following
\begin{eqnarray}
2\pi \Delta E(x^\pm) &=&
 \int_{-1}^1 dr  \sqrt{ k(r)^2+ m^2} \frac{\partial}{\partial r }\left[
 \delta_{Z_2} + \delta_{\bar Z_2} - ( \delta_{\theta} + \delta_{\eta} + \delta_{\bar\theta}
 + \delta_{\bar\eta}) \right],  \\ \nonumber
&=&
  \int_{-1}^1 dr  \sqrt{ k(r)^2+ m^2}
\frac{\partial}{\partial r} \left[ -2G(r; x^\pm) + 2 G(1/r; x^\pm) \right. \\ \nonumber
& & \left. \qquad\qquad - 2    G(1/r, x^\pm) + G(1/r, x^\pm)
\right] ,\\ \nonumber
&=&0.
\end{eqnarray}
We have changed the variable of integration from the momentum of the
plane wave to $r$.  Thus the one loop correction to the energy of the
dyonic magnon vanishes.  Since the giant magnon can be obtained as
a smooth limit of the dyonic giant magnon, this also implies that the one
loop correction to the energy of the giant magnon vanishes.
From the discussion in section 2.2 we see that
this implies the coefficient of one loop correction $g_0$  in (\ref{strong-exp}) vanishes.
Thus the relation
$4g = R^2/\pi\alpha'$
is true to one loop in the strong coupling expansion.
Applying this result to the dyonic giant magnon dispersion relation in
(\ref{dyon-sdisp}) we see that possible one loop corrections to the relation does not exist
and the dispersion is one loop exact. This must be the case  since  the dispersion
relation arises as a result of a BPS condition.

\subsection{One loop scattering phase}

Using (\ref{1-lphase}) the one loop correction to the
scattering phase of two dyonic magnons can be found, but then using
the relation (\ref{rel-phase})   and the explicit
values of the phase shifts evaluated in (\ref{end-phase})
we can evaluate the one loop contribution to the
phase factor in $\sigma(x^\pm, y^\pm)$.  This is given by
\begin{eqnarray} \label{dpcorr}
& & 2\theta_1(x^{\pm},y^{\pm}) \\ \nonumber&=&
\frac{1}{4\pi}\int_{-\infty}^\infty dr \left (  \left[
\frac{\partial \delta_{Z_2}(r, x^\pm) }{\partial r}  \delta_{Z_2} (r, y^\pm)
- \frac{\partial \delta_{\theta}( r,x^\pm)}{\partial r}  \delta_{\theta}(r,y^\pm)
-\frac{\partial \delta_{\eta} (r,x^\pm)}{\partial r}  \delta_{\eta}(r, y^\pm)\right] \right. \\ \nonumber
& & \left. +  \left[
\frac{\partial  \delta_{\bar Z_2} (r, x^\pm) }{\partial r} \delta_{\bar Z_2}( r, y^\pm)
- \frac{\partial \delta_{\bar \theta} ( r,x^\pm)}{\partial r}
 \delta_{\bar \theta}(r,y^\pm)
-\frac{\partial \delta_{\bar \eta}( r,x^\pm)}{\partial r}
\delta_{\bar \eta}(r, y^\pm)\right]\right)
\\ \nonumber
& & - (x^\pm \leftrightarrow y^\pm  ), \\ \nonumber
&= &
 \frac{1}{2\pi}
\left[ \int_{-1}^{+1} ~ dr ~ \frac{\p G(r,x^{\pm})}{\p r}
 G(r,y^{\pm}) ~ + \int_{-1}^{+1}~dr ~ \frac{\p G(\frac{1}{r},x^{\pm})}{\p r}
G(\frac{1}{r},y^{\pm})\right] \\ \nonumber
& &  - (x^\pm \leftrightarrow y^\pm  ),
\end{eqnarray}
where
\begin{equation}
\label{val-g}
G(r,x^{\pm})=- \left( \frac{1}{i} \ln\left(\frac{r-x^+}{r-x^-}\right) - \frac{1}{2i}
\ln\frac{x^+}{x^-} \right).
\end{equation}
 Hence
\ben \label{der}
\frac{\p G(r,x^{\pm})}{\p r}&=& i\left(\frac{1}{r-x^+}-\frac{1}{r-x^-}\right), \nonumber \\
\frac{\p G(\frac{1}{r},x^{\pm})}{\p r} &=&
 i\left(\frac{1}{r-\frac{1}{ x^+}}-\frac{1}{r-\frac{1}{x^-}}\right).
\een
Using the above
 expressions,  the one loop correction to the dressing phase can be arranged as
\be
2 \theta_1(x^{\pm},y^{\pm})=\chi_1(x^+,y^+)-\chi_1(x^+,y^-)-\chi_1(x^-,y^+)+
\chi_1(x^-,y^-),
\ee
where
\ben
\label{def-chi1}
\chi_1(x,y)&=&-\frac{1}{2\pi}\left[\int_{-1}^{+1}\frac{dr}{r-x}
\left( \ln(r-y) - \frac{1}{2} \ln y\right)  \right. \\ \nonumber
& & \left. +\int_{-1}^{+1}~\frac{dr}{r-\frac{1}{x}}
\left(\ln(1/r-y) - \frac{1}{2} \log y\right) - (x\leftrightarrow y) \right] \nonumber, \\
 &\equiv & -\frac{1}{2\pi}\left[I_1(x,y) - I_1(y, x) + I_2 (x, y)-I_2(y, x) \right] .
\een
The above integrals can be integrated using
$$-\int dt~\frac{\ln(1-t)}{t}={\rm{Li}}_2(t),$$ where ${\rm{Li}}_2(t)$ is a dilogarithmic function of argument $t$.
 We obtain the result
\ben \label{int}
I_1(x, y) &=&\left( \ln(x-y)- \frac{1}{2}\ln(y)\right)
 \ln{\left(\frac{x-1}{x+1}\right)}+\left[{\rm{Li}}_2{\left(\frac{x+1}{x-y}\right)}- {\rm{Li}}_2{\left(\frac{x-1}{x-y}\right)} \right],\nonumber \\
I_2(x,y) &=&\left(  \ln{\left(1-\frac{y}{x}\right)} - \frac{1}{2}
\ln(y) \right)\ln{\left(\frac{1-x}{1+x}\right)}\nonumber \\
&& +
\left[{\rm{Li}}_2{\left[(x+1)\frac{y}{y-x}\right]}-
{\rm{Li}}_2{\left[(x-1)\frac{y}{x-y}\right]}\right] .
\een

\section{Unitarity check on the crossing relations}

S-matrices obey crossing relations. These are conditions which
are obtained when one of the Hilbert space the S-matrix acts on is replaced
by its anti-particle. Formally the conditions can be written as the following
\begin{eqnarray}
\label{def-crossing}
 {\cal C}^{-1}\otimes I {\cal S}_{12}^{T_1} (- p_1, p_2) {\cal C}
\otimes I {\cal S}_{12} (p_1, p_2) = I, \\ \nonumber
I\otimes {\cal C}^{-1}{\cal S}_{12}^{T_2}(p_1, - p_2) I\otimes {\cal C} {\cal S}_{12} (p_1, p_2) = I .
\end{eqnarray}
Where ${\cal C}$ is the charge conjugation operation, $T_1, T_2$ refer to the transpose
operations on the first and second Hilbert space respectively.
As we have seen symmetries constrain the S-matrix to a form given by
\begin{equation}
{\cal S}_{12}(p_1, p_2)  = S_0(p_1, p_2) \hat{\cal S}_{12}(p_1, p_2),
\end{equation}
where $\hat{\cal S}_{12}(p_1, p_2)$ is completely determined by symmetries and
$S_0(p_1, p_2)$  is the scalar function  which cannot be determined
by symmetries alone. It satisfies the unitarity condition
\begin{equation}
\label{unit}
S_0(p_1, p_2) S_0(p_2, p_1) = 1.
\end{equation}
Then substituting the above form for the S-matrix in (\ref{def-crossing})
one obtains the following conditions on the scalar function
\begin{eqnarray}
\label{cross-1}
S_0(-p_1, p_2) S_0(p_1, p_2) = f(p_1, p_2), \\ \nonumber
S_0( p_1, -p_2) S_0(p_1, p_2) = g(p_1, p_2).
\end{eqnarray}
These are part of the consistency conditions which are necessary
for the  S-matrix to satisfy the crossing symmetry relations in
(\ref{def-crossing}). The  unitarity condition in
(\ref{unit})  then implies the following constraint
on the function $f$.
\begin{equation}
\label{u-cond}
f(p_1, p_2) g(p_2, p_1) =1.
\end{equation}

The function $f(p_1, p_2)$ can be evaluated if
one can implement the transformation to the anti-particle by means of
the antipode operation as it was done for the
case of $AdS_5\times S^5$ in \cite{Janik:2006dc}. The validity
of the function $f(p_1, p_2)$  for this case  was tested
to one loop in the strong coupling expansion in \cite{Arutyunov:2006iu}.
The energy and momentum of the anti-particle is equal in magnitude but
 opposite in sign to the particle.
 Translating this to the spectral parameters $x^\pm$ it can be seen
that the spectral parameters of the anti-particle is
 related to the particle by the following
 \begin{equation}
 \bar x^\pm = \frac{1}{x^\pm},
 \end{equation}
 where the bar denotes the parameters for the anti-particle.
 From (\ref{xpm}) it is easy to see that the  above transformation reverses the
 sign of the momentum. To see that it also  changes the sign of the
 energy given by $C_1 + C_2 = 2gc -1$, first use (\ref{solxpm}) to
 show
 \begin{equation}
 \frac{1}{x^-} = \frac{i}{2g( e^{-ip} -1) }
  \left( 1- \sqrt{ 1 + 16 g^2 \sin^2\frac{p}{2} } \right).
\end{equation}
Using this one can evaluate
\begin{eqnarray}
\bar C_1 + \bar C_2 &=& 2g \bar c -1, \\ \nonumber
&=& -i2g\left( \frac{1}{x^+} - \frac{1}{x^-} \right) -1, \\ \nonumber
&=& -   \sqrt{ 1 + 16 g^2 \sin^2\frac{p}{2} }.
\end{eqnarray}
However it is shown in the appendix C, the antipode operation
for the $SU(1|1)$ invariant S-matrix does not implement the
above transformation  on the spectral parameters.
Instead, it changes $x^+ \leftrightarrow x^-$. This certainly reverses
the sign of the momentum, but does not change the sign of the energy.
Thus we are unable to determine the function $f(p_1, p_2)$
by the means of the antipode operation. However we can still test the
whether the unitarity condition on the function $f$ given in (\ref{u-cond})
holds. In terms of spectral parameters this condition is given by
\begin{equation}
\label{unitarity}
f(x, y) g( y, \frac{1}{x}) =1.
\end{equation}
Here  and in the rest of this section $x, y$ refer to the variables $x^\pm, y^\pm$
respectively.
Note the inversion of the variable in the second term, this is due to the
fact the crossing conditions are formulated in the universal cover of the
spectral parameter plane \cite{Janik:2006dc}.
Translating the equations in (\ref{cross-1}) in terms of the spectral parameters
we have
\begin{equation}
\label{crossing}
S_0(x, y) S_0(\frac{1}{x}, y) = f(x,y), \qquad
S_0(x,y) S_0(x, \frac{1}{y}) = g(x,y).
\end{equation}
From (\ref{unitarity}) and the
form for $S_0$ given in (\ref{phase}) we see that we obtain the following constraint
on the dressing factor $\sigma(x, y)= \exp( i \theta(x, y))$
\begin{equation}
\label{unit-constraint}
\ln ( \frac{y^+}{y^-} ) + i\left[  \theta(x, y) +  \theta\left(\frac{1}{x}, y\right)\right] =
-  \ln( \frac{x^-}{x^+}) - i \left[\theta\left( y, \frac{1}{x} \right) +
\theta\left(\frac{1}{y}, \frac{1}{x}\right)\right].
\end{equation}
The above equation is obtained by
taking the logarithm of
(\ref{unitarity}) and substituting  for $f(x, y), g(x, y)$ from (\ref{crossing}).
We then use the antisymmetry property of $\theta (x,y)$ to arrive
 the following constraint on the dressing factor
\be \label{unit-constraint1}
\ln{y^{+}\over y^{-}}+i\theta(x,y)-
\ln{x^{+}\over x^{-}}-i\theta\left({1\over x},{1\over y}\right)=0.
\ee

We now show that   phase $\theta_0$ determined up to
one loop in the sigma model coupling satisfies the
constraint (\ref{unit-constraint1}).
As we have seen the dressing phase $\theta(x, y)$ is expanded as follows.
\begin{eqnarray}
\label{def-theta}
 \theta(x, y) &=& g\theta_0 (x,y)+  \theta_1 (x,y) + \cdots.
\end{eqnarray}
Using the form for
$\theta_0(x, y)$ in
 (\ref{lead-phase1}) and (\ref{lead-phase}), one can
  deduce the following relation for $\theta_0$, which is
\ben \label{31a}
\theta_0\left({1\over x},{1\over y}\right)&=& \theta_0(x,y)-\ln\left({{x^+}\over{x^-}}\right)\left(y^- +{1\over y^-}-y^+
- {1\over y^+}\right) \nonumber \\
&& +\ln\left({{y^+}\over{y^-}}\right)\left(x^- +{1\over x^-}-x^+
- {1\over x^+}\right).
\een
Now substituting the constraint satisfied by the
spectral parameters in (\ref{xpm-1}) we get
\be \label{33}
g\theta_0\left({1\over x},{1\over y}\right)= g\theta_0(x,y)+i\ln\left({{x^+}\over{x^-}}\right)-i\ln\left({{y^+}\over{y^-}}\right).
\ee
Let us now examine the one loop correction to the phase $\theta_1(x,y)$ given
in (\ref{dpcorr}). From (\ref{val-g}), we see that $G(r,{1\over x^{\pm}})=G({1\over r},x^{\pm})$. Using this property, it is easy to see that the
 one loop dressing phase given in  \refb{dpcorr} satisfies the relation
\be \label{31}
\theta_1\left({1\over x},{1\over y}\right)=\theta_1(x,y).
\ee
Combining the results of
(\ref{33}) and (\ref{31}), we
see that the constraint on the dressing factor
 \refb{unit-constraint1} is satisfied.
 Note that we have not resorted to any expansion of the spectral parameters in
 terms of the momentum and the coupling to verify this constraint
 \footnote{We have verified that  the
 dressing phase obtained for ${\cal N}=4$ SYM at one loop
 satisfies a similar constraint.}.

\section{Conclusions}

We have seen that $SU(1|1)$ symmetries constrain the S-matrix of the magnon
excitations with polarizations in $AdS_3\times S^3$ up to a phase.
We have  determined the phase in the sigma model expansion to
one loop. We then show that that phase satisfies the constraint of unitarity
implied by crossing symmetry.
Using the semi-classical methods we also  showed that the one loop correction to the
dispersion relation vanishes at strong coupling.

In \cite{Babichenko:2009dk}  a proposal for the quantum Bethe equations which
are valid at all values of coupling was made from the symmetries of the
coset description of the
Green-Schwarz string on $AdS_3\times S^3$.
The Bethe equations were undetermined up to a phase, our work provide
the value of this phase
to one loop at strong coupling.
To make this more explicit one should
 identify the
magnons described here as a sub-sector of the full theory and show that Bethe equations proposed in  \cite{Babichenko:2009dk} in this sub-sector reduce to the
Bethe equations obtained from the S-matrix given in this paper.
The S-matrix  in this paper is invariant under one of $SU(1|1)$ symmetries
using the ordinary co-product while the second $SU(1|1)$ is realized as
a non-trivial co-product. Using this fact it is perhaps possible to implement
crossing symmetry  with some modifications of the approach in
 \cite{Janik:2006dc} and determine the function
$f(x, y), g(x, y) $ in the crossing equations given in (\ref{crossing}).
This will enable one to make more consistency checks on the phase
$\theta(x, y)$ and perhaps determine it completely.

\acknowledgments

B. Sahoo thanks the Centre for High Energy Physics, IISc, Bangalore
 for hospitality during part of this work.
We thank Diptiman Sen for a discussion on spin chains and Yang-Baxter relations
and Sachin Vaidya for a discussion on non-trivial co-products.

\appendix

\section{The extended $SU(1|1)\times SU(1|1)$ algebra}

It was shown in \cite{David:2008yk}, that that magnons in the D1-D5 system are BPS states  in  the extended  $SU(1|1)\times SU(1|1)$ algebra given in (\ref{ext-algeb}).
To make the paper self contained we derive their dispersion relation from the extended
algebra. The derivation presented here is simplified compared to the
one in \cite{David:2008yk}.

We write down the simplest irreducible representation
of the extended algebra (\ref{ext-algeb}) relevant for the
magnons. The ground state is
$|\phi_p \rangle \otimes | 0\rangle$.
The action of the charges on the ground state is given by
\begin{eqnarray}
\label{map1}
Q_1|\phi_p \rangle |\otimes | 0\rangle = a(p) |\psi_p \rangle\otimes| | 0\rangle,
&\qquad& S_1   |\phi_p \rangle \otimes | 0\rangle = 0 =0, \\ \nonumber
Q_2 |\phi_p \rangle \otimes | 0\rangle = 0, &\qquad&
S_2 |\phi_p \rangle \otimes | 0\rangle = b'(p) |\phi_p\rangle \otimes |\psi^- \rangle.
\end{eqnarray}
Now from analysis of first order perturbation theory in the $\ZZZ_2$ twisted operator
in the symmetric product of the D1-D5 system the following identification
of the excited states can be inferred (See eq (2.9) of
 \cite{David:2008yk}, use the map in ( \ref{unclutter} ) to
 translate the notations for the charges )
\begin{equation}
\label{identi}
|\psi_p \rangle\otimes| | 0\rangle = |\phi_p\rangle \otimes |\psi^- \rangle.
\end{equation}
This identification cuts down the number of states in the representation to
2, which is the right number for a half BPS state in the extended
algebra (\ref{ext-algeb}).
Due to this identification we can restrict ourselves to the single excited state
$|\psi_p\rangle\otimes|0\rangle$ and we can rewrite the last equation in
(\ref{map1}) as
\begin{equation}
S_2 |\phi_p \rangle \otimes | 0\rangle = b' (p)|\psi_p\rangle \otimes |0\rangle.
\end{equation}
 The action of the charges  on the excited state are given by
\begin{eqnarray}\label{map2}
Q_1 |\psi_p\rangle\otimes |0\rangle =0,
&\qquad&
S_1 |\psi_p \rangle\otimes |0\rangle = b (p)|\phi_p\rangle \otimes 0\rangle,
\\ \nonumber
Q_2|\psi_p\rangle \otimes|0\rangle = a'(p) |\phi_p\rangle \otimes |0\rangle,
&\qquad&
S_2  |\phi\psi^+ \rangle\otimes |0\rangle = 0.
\end{eqnarray}
It is easy to see that the above actions of the charges on states are
consistent  with the nilpotent relations of the
algebra that is $Q_1^2 = Q_2^2= S_1^2= S_2^2 =0$. From
(\ref{map1}) and (\ref{map2}) we also  see that for the
states in this multiplet,  the action of $Q_1$ is proportional to the action of $S_2$
and the action of $S_1$ is proportional to the action of $Q_2$. We can write
this as
\begin{equation}
\label{charge-eq}
b'(p) Q_1= a(p) S_2, \qquad  a'(p) S_1 = b(p) Q_2.
\end{equation}
We can now evaluate the central charges on the various states of this
irreducible multiplet. From the algebra is is clear that the
various central charge of the entire multiplet must be same.
\begin{eqnarray}
\{ Q_1, Q_2 \}  |\phi_p\rangle \otimes 0\rangle =
a(p)a' (p) |\phi_p\rangle \otimes 0\rangle,
\\ \nonumber
\{ Q_1, Q_2 \} |\psi_p \rangle\otimes |0\rangle =
a(p)a' (p)|\psi_p\rangle \otimes 0\rangle.
\end{eqnarray}
From the definition of the extended algebra in (\ref{ext-algeb}) we see that
this results in the following value of  the central charge for a single magnon
with momentum $p$
\begin{equation}
C_3 - i C_4 = a(p)a'(p).
\end{equation}
Similarly  we can show that the other central charges are
\begin{equation}
C_3 + i C_4 = b(p)b'(p), \qquad C_1 =a(p)b(p), \qquad C_2 = a'(p) b'(p).
\end{equation}
We can also show the relation $\{Q_1, S_2\} =\{Q_2, S_1\} =0$
holds on the states of the short multiplet using the
relations in \eq{map1}, \eq{map2}.

We now wish to find the dependence of the central charges
$C_3 \pm i C_4$ on the momentum of the magnon.
For this it is useful to examine the action of these central charges on a
two magnon state.  From the spin chain description of the magnons introduced
in \cite{David:2008yk} it can be seen that the action of these central charges on a
2 magnon state is given by
\begin{eqnarray}
\nonumber (C_3 -i C_ 4)
|\phi_{p_1}\phi_{p_2}\rangle \otimes |0\rangle =
[ a(p_1) a'(p_1) \exp(- ip_2) + a(p_2) a'(p_2)]
 |\phi_{p_1}\phi_{p_2}\rangle
 \otimes |0\rangle. \\
\end{eqnarray}
The reason the phase $\exp(- ip_2) $ occurs in the first term is because the
the central charge $C_3 \pm i C_4$ changes the length the states,
which translates to a insertion of additional   momentum
factors \cite{David:2008yk}.  Another way of thinking of this action is that
that the tensor product of the two magnons involves a non-trivial
co-product \cite{Gomez:2006va,Plefka:2006ze, Arutyunov:2006yd}. Therefore on multi-magnon states the
action of the central charge is given by
\begin{equation}
C_3 - i C_4 |\phi_{p_1}\phi_{p_2} \cdots \phi_{p_j} \rangle\otimes |0\rangle
= \sum_{k=1}^j a(p_j) a'(p_j) \prod_{l= k+1}^j \exp( -i p_l)
|\phi_{p_1}\phi_{p_2} \cdots \phi_{p_j} \rangle\otimes |0\rangle.
\end{equation}
Now on physical states the total central charge should vanish
and the extended algebra  (\ref{ext-algeb}) should reduce to the usual
$SU(1|1)\times SU(1|1)$ algebra. This fixes the form of $a(p_j)a'(p_j)$ to
be
\begin{equation}
a(p_j) a'(p_j)  = \alpha( \exp( -i p_j) -1),
\end{equation}
where $\alpha$ is a constant independent of the momentum.
Thus the total central charge on the multi-magnon
state of $j$ magnons is given by
\begin{equation}
C_3 - iC_4 = \alpha\sum_{k=1}^j ( \exp( -i p_k -1) \prod_{l=k+1}^j \exp( -i p_l))
= \alpha( \prod_{k=1}^j \exp( -i p_k) -1).
\end{equation}
Therefore on states which satisfy this physical state
condition $\sum_{k=1}^j p_k =0$,
the additional central charge $C_3 - iC_4$ vanishes.
A similar argument for the central charge $C_3+iC_4$ results in the equation
\begin{equation}
b(p_j)b'(p_j) = \alpha^* ( \exp( i p_j) -1).
\end{equation}
We are now in a position to derive the dispersion relation for a single
magnon with momentum $p$. We have
\begin{eqnarray}
\label{val-c1}
a(p)a'(p) = \alpha( e^{-i p} -1), \qquad
b(p)b'(p) = \alpha^*( e^{-ip} - 1) , \\ \nonumber
C_1 +C_2 = a(p)b(p) - a'(p)b'(p), \qquad C_1 - C_2 = 1 = a(p)b(p) - a'(p)b'(p).
\end{eqnarray}
The reason $C_1-C_2=1$ is because, we are dealing with a single magnon.
Note that the values of the central charges in (\ref{val-c1})
satisfies the BPS condition (\ref{bps-cond}).
Now as a result of these equations we have
\begin{equation}
C_1 + C_2 = \sqrt{ 1+ 16 |\alpha|^2  \sin^2 ( \frac{p}{2} ) }.
\end{equation}
This completes our derivation of the magnon dispersion relation.

\subsection{Invariance of the S-matrix}

Due to the fact that the S-matrix is constructed out of the Casimirs of the first
$SU(1|1)$, that is the algebra generated by the charges $Q_1, S_2, C_1$
it is invariant under the trivial co-product of this algebra as is seen
in (\ref{triv-cop}).
From  the relations in (\ref{charge-eq}) which hold on states of the short multiplet,
we can show that  the invariance with respect to the second
$SU(1|1)$ generators is of the following form
\begin{eqnarray}
 [ \frac{a(p)^{(1)} }{ b'(p)^{(1)} } S_{2}^{(1)} \otimes 1 + (-1)^F \otimes
\frac{a(p)^{(2)}} { b'(p)^{(2)} }S_2^{(1)} , {\cal S}_{12} ] =0, \\ \nonumber
[ \frac{b(p)^{(1)} }{ a'(p)^{(1)} }Q_{2}^{(1)} \otimes 1 + (-1)^F \otimes
\frac{b(p)^{(2)}} { a'(p)^{(2)} }S_2^{(1)} , {\cal S}_{12} ] =0.
\end{eqnarray}
Here the superscripts refer to the two states in the tensor product.
The above relations can also be written in terms of the central charges
$C_3 \pm i C_4$ and $C_1, C_2$ as the follows: let us define
\begin{equation}
 C_3 + iC_4 =  { C}, \qquad C_3 - iC_4 = \bar { C}.
\end{equation}
Re-writing the above equations in terms of the central charges we obtain
\begin{eqnarray}
\label{inv-smat}
 [{{ C}^{(1)} }  S_{2}^{(1)} \otimes C_2^{(2)} + (-1)^F C_2^{(1)}
\otimes {{ C}^{(2)} } S_{2}^{(2)}, {\cal S}_{12} ] =0, \\ \nonumber
 [{\bar{ C}^{(1)} } Q_{2}^{(1)} \otimes C_2^{(2)} + (-1)^F C_2^{(1)}
\otimes {{\cal C}^{(2)} } Q_{2}^{(2)}, {\cal S}_{12} ] =0.
\end{eqnarray}
Note that the factors which arise in the invariance of the S-matrix in (\ref{inv-smat}) are momentum
dependent factors involving the central charges.
Thus the invariance of the S-matrix under the second $SU(1|1)$
is realized under the above non-trivial co-product.

\section{Phase shifts by the dressing method} \label{dressappendix}

As discussed in section \ref{dressmain}, the dressing phase method devised in
 \cite{Spradlin:2006wk} is a powerful method to obtain a $N+1$-soliton
solution from a $N$ soliton solution.  We can use this method to find the phase shifts suffered by a plane wave scattering in
an $N$-DGM background taking the plane wave limit on one of the dyonic giant magnon.
We will first
 outline the main points of the dressing method ,
 for details see \cite{Spradlin:2006wk}, see also \cite{Kalousios:2008gz} for a recent
application of this method to the case of $S^3$ where the phase shift and the time delay for the 
scattering of $N$ magnons has been evaluated:

\begin{enumerate}
\item
We begin by introducing an auxiliary complex parameter $x$ and a set of three matrices ($\psi$($x$), $A$ and $B$) which satisfies the system of equations
 \be \label{seom}
 i\bar{\partial}\psi={{A\psi}\over{1+x}}, \quad i{\partial}\psi={{B\psi}\over{1-x}},
 \ee
with $A$ and $B$ independent of $x$, the spectral parameter.
It is then easy to verify that $\psi$($0$) satisfies the equation of motion for  $g$ given in \refb{eom}. We also
need to impose the $SU(2)$ condition on the matrix $\psi(x)$.
\item
Now consider the  transformation
\ben \label{transf}
\psi &\rightarrow & \psi^{\prime}=\chi\psi \nonumber \\
A &\rightarrow & A^{\prime}=\chi A \chi^{-1}+i\left(1+x\right)\bar{\partial}\chi \chi^{-1}, \nonumber \\
B &\rightarrow & B^{\prime}=\chi B \chi^{-1}+i\left(1-x\right)\bar{\partial}\chi \chi^{-1}.
\een
If we can choose $\chi$ in such a way that the new $A^{\prime}$ and $B^{\prime}$ are independent of $\lambda$, then the set ($\psi^{\prime}$($\lambda$), $A^{\prime}$ and $B^{\prime}$) will be a new solution to \refb{seom} and hence provide a new solution $g^{\prime}=\psi^{\prime}(0)$to the principal chiral model \refb{eom}. The dressing function $\chi$ is completely fixed by the above requirements to be \cite{Spradlin:2006wk}
\be \label{1.3}
\chi_N(x)=\left(1+\frac{x_N-\bar{x}_N}{x-x_N}P_{N}\right)\sqrt{\frac{x_N}{\bar{x}_N}}.
\ee
Where
\be \label{1.4}
P_N=\frac{\psi_{N-1}(\bar{x}_N)ee^{\dag}\psi^{\dag}_{N-1}(\bar{x}_N)}{e^{\dag}\psi^{\dag}_{N-1}(\bar{x}_N)\psi(\bar{x}_N)e},
\ee
and $e$ is the constant column  vector  $(1,1)$. We have put a subscript $N$ in $\chi$ to signify that one can successively dress starting from a solution and $N$ refers to the
number of the step in the iteration.
\item
 The vacuum solution for $\psi$ is given by
\be \label{1.2}
\psi_0= \left(
\begin{array}{cc}
e^{iZ(x)} & 0 \\
0 & e^{-iZ(x)} \\
\end{array}
\right), \quad \quad \mbox{where} ~
Z(x)=\frac{z_{-}}{x-1}+\frac{z_{+}}{x+1}, \quad z_{\pm}=\frac{1}{2}(\sigma \pm t).
\ee
$g_0=\psi_0(0)$ will then be a point like string rotating around the equator of the S$^3$.
\item
We now dress the solution using the dressing factor \refb{1.3} on the
vacuum solution   to get multi-solition solutions. Given a $N$-soliton solution
multiplying it by the dressing factor \refb{1.3} we can dress it to a $N+1$ soliton
solution.
\be \label{1.5}
\psi_{N+1}=\chi_{N+1}\psi_N.
\ee
Thus by starting from the vacuum solution $\psi_0$ we can dress it to get 1-soliton,
2-soliton and so on. At the end we will identify the parameters, appearing in the dressing factor $\chi_N$, with the spectral parameters of the solution as
$x_N \leftrightarrow x_{N}^+, \bar{x}_{N}\leftrightarrow x_{N}^{-}$ the spectral
parameters of the $N$ soliton solution.
\end{enumerate}

We can now apply the
dressing method to
calculate the phase shift suffered by a plane wave  scattering off
the background of the $N$-DGM solution. To obtain the phase shift
 we have to dress $\psi_N$ to get $\psi_{N+1}$
 and take the plane wave limit  on the last $N+1$th magnon.
 This essentially means that  we need to take the spectral
 parameter of the last $N+1$ magnon to be real.
Let  $x_{N+1} = r\exp( i\frac{q}{2})$, it will turn out we will need to examine the
solutions
 up to terms linear  in $q$.  We have
\be \label{1.6}
\psi_{N+1}=\chi_{N+1}\psi_{N}=\left(1+\frac{x_{N+1}-\bar{x}_{N+1}}{x-x_{N+1}}P_{N+1}\right)\sqrt{\frac{x_{N+1}}{\bar{x}_{N+1}}}\psi_N, \ee
let us define $g_N=\psi_N(0)$, then it  follows from the above equation
\ben \label{1.7}
g_{N+1}&=& \sqrt{\frac{x_{N+1}}{\bar{x}_{N+1}}}g_N-\frac{x_{N+1}-\bar{x}_{N+1}}{x_{N+1}}P_{N+1}\sqrt{\frac{x_{N+1}}{\bar{x}_{N+1}}}g_N, \nonumber \\
&=& e^{i\frac{q}{2}}g_N-2i (\sin {\frac{q}{2}})P_{N+1}g_N.
\een
Now we need the asymptotic behavior i.e $\sigma \to \pm \infty$ up to terms linear in q. So we have \footnote{We use the shorthand $P_{N+1}(r,\pm \infty)$ to denote $P_{N+1}(x_N=\bar{x}_N=r, \sigma \to \pm \infty)$ and $g_N(\pm \infty)$ to denote
$g_N(\sigma \to \pm \infty)$}
\ben \label{1.8}
&& g_{N+1}(\pm \infty)=g_{N}(\pm \infty)+i (\sin {\frac{q}{2}})g_{N}(\pm \infty)- 2i (\sin {\frac{q}{2}})P_{N+1}(r,\pm \infty)g_N(\pm \infty), \nonumber \\
\Rightarrow && \delta g_{N}(\pm \infty) \equiv g_{N+1}(\pm \infty)-g_{N}(\pm \infty)=i (\sin {\frac{q}{2}})(1-2P_{N+1}(r,\pm \infty))g_N(\pm \infty). \nonumber \\
\een
So in order to determine $\delta g_{N}(\pm \infty)$, we need to know $g_N(\pm \infty)$ and $P_{N+1}(r,\pm \infty)$.  To determine $g_N(\pm \infty)$ we first note that
\ben \label{1.9}
P_i(x_i \neq \bar{x}_i, \sigma \to \infty)&=&\left( \begin{array}{cc}
0 & 0 \\
0 & 1 \\
\end{array}\right), \nonumber \\
P_i(x_i \neq \bar{x}_i, \sigma \to -\infty)&=& \left(\begin{array}{cc}
1 & 0 \\
0 & 0 \\
\end{array}\right).
\een
Using the dressing prescription we get
\be \label{1.10}
\psi_N(x,\infty)=\sqrt{\frac{x_1 x_2 \cdots x_N}{\bar{x}_1
\bar{x}_2 \cdots \bar{x}_N}}\left(\begin{array}{cc}
1 & 0 \\
0 \ & {\prod_{i=1}^{N}\, \left(\frac{x-\bar{x}_i}{x-x_i}\right)} \\
\end{array}\right)\left(\begin{array}{cc}
e^{iZ(x)} & 0 \\
0 & e^{-iZ(x)} \\
\end{array}\right).
\ee
After substituting  $x_i \equiv r_i \, e^{\frac{ip_i}{2}}$, we obtain
\be \label{1.11}
\psi_N(x,\infty)=e^{\frac{iP}{2}}\left(\begin{array}{cc}
e^{iZ(x)} & 0 \\
0 & \prod_{i=1}^{N}\, \left(\frac{x-\bar{x}_i}{x-x_i}\right)e^{-iZ(x)} \\
\end{array}\right), \quad P\equiv\sum_{i=1}^{N}p_i,
\ee
and similarly
\be \label{1.12}
\psi_N(x,-\infty)=e^{\frac{iP}{2}}\left(\begin{array}{cc}
\prod_{i=1}^{N}\, \left(\frac{x-\bar{x}_i}{x-x_i}\right) e^{iZ(x)} & 0 \\
0 & e^{-iZ(x)} \\
\end{array}\right).
\ee
As a consequence  of these equations we get
\be \label{1.13}
g_{N}(-\infty)= \left(\begin{array}{cc}
e^{i(t-\frac{P}{2})} & 0 \\
0 & e^{-i(t-\frac{P}{2})} \\
\end{array}\right), \quad g_{N}(\infty)= \left(\begin{array}{cc}
e^{i(t+\frac{P}{2})} & 0 \\
0 & e^{-i(t+\frac{P}{2})} \\
\end{array}\right).
\ee
Now our
next task is to determine $P_{N+1}(r,\pm \infty)$ for which we shall use \refb{1.4} .
This requires the expression for  $\psi_N(r,\pm \infty)$, which we read off from
(\ref{1.11}) and (\ref{1.12}). The result is given by
\ben \label{1.14}
P_{N+1}(r,\infty)&=&\frac{1}{2}\left(\begin{array}{cc}
1 & \prod_{i=1}^{N}\bar{\Sigma}_i \ e^{2iZ(r)} \\
\prod_{i=1}^{N}{\Sigma}_i \ e^{-2iZ(r)} & 1 \\
\end{array}\right), \\ \nonumber
P_{N+1}(r,-\infty)&=&\frac{1}{2}\left(\begin{array}{cc}
1 & \prod_{i=1}^{N}{\Sigma}_i \ e^{2iZ(r)} \\
\prod_{i=1}^{N}\bar{\Sigma}_i \ e^{-2iZ(r)} & 1 \\
\end{array}\right), \quad \Sigma_i \equiv \frac{r-\bar{x}_i}{r-x_i}.
\een
Finally using (\ref{1.8}), (\ref{1.13}), and (\ref{1.14}), we obtain
\ben \label{1.15}
\delta g_N(+\infty)&=& -i \sin{\frac{q}{2}}\left(\begin{array}{cc}
0 & \prod_{i=1}^{N}\bar{\Sigma}_i \ e^{i(2Z(r)-t-\frac{P}{2})} \\
\prod_{i=1}^{N}{\Sigma}_i \ e^{-i(2Z(r)-t-\frac{P}{2})} & 0 \\
\end{array}\right), \nonumber \\
\delta g_N(-\infty)&=& -i \sin{\frac{q}{2}}\left(\begin{array}{cc}
0 & \prod_{i=1}^{N}{\Sigma}_i \ e^{i(2Z(r)-t+\frac{P}{2})} \\
\prod_{i=1}^{N}\bar{\Sigma}_i \ e^{-i(2Z(r)-t+\frac{P}{2})} & 0 \\
\end{array}\right).
\een
From \refb{1.2} one can check that
\be \label{1.161}
2Z(r)-t=\omega t-kx, \quad  \omega=\frac{1+r^2}{1-r^2},\quad  k=\frac{2r}{1-r^2}.
\ee
Thus the perturbation is a plane wave of wave number $k$ and frequency $\omega=\sqrt{1+k^2}$. Now from \refb{1.15}, we get
\ben \label{1.16}
\delta Z_1(\pm \infty)&=&0 \nonumber \\
\delta Z_2(\infty)&=& \sin {\frac{q}{2}} \prod_{i=1}^{N} \bar{\Sigma}_i \ e^{i(\omega t - kx -\frac{P}{2})} \nonumber \\
\delta Z_2(-\infty)&=& \sin {\frac{q}{2}} \prod_{i=1}^{N} {\Sigma}_i \ e^{i(\omega t - kx +\frac{P}{2})}
\een
Thus the phase shifts for the perturbation associated with the S$^3$ fluctuations around a $N$-DGM background are
\ben \label{1.17a}
\delta_{Z_1} &\equiv& -i\ln(\delta Z_1(\infty))+i\ln(\delta Z_1(-\infty))=0, \\ \nonumber
\delta_{Z_2} &\equiv& -i\ln(\delta Z_2(\infty))+i\ln(\delta Z_2(-\infty))=2i \ln(\prod_{i=1}^{N}\ \Sigma_i)\ -P, \\ \nonumber
&=& - i \sum_{i=1}^N \left[ 2\ln\left( \frac{r - x_i^+}{r- x_i^-} \right)  -
\ln\left( \frac{x_i^+}{x_i^-}\right) \right].
\een
Here  we have used the fact that $\bar{\Sigma}_i=\frac{1}{\Sigma_i}$, which can be easily verified
from \refb{1.14}.
To obtain the last line in (\ref{1.17a}) we have reinstated the definition of the
spectral parameters $x^+_i = x_i$ and $x^- = \bar x_i$.
One can also show that  the phase shifts for the complex conjugate fields
are given by
\begin{eqnarray}
\label{1.18a}
 \delta_{\bar Z_1}(1/r)  &\equiv& -i\ln(\delta \bar{Z_1}(\infty))+i\ln(\delta \bar{Z_1}(-\infty))= 0, \\ \nonumber
\delta_{\bar Z_2}(1/r)   &\equiv& -i\ln(\delta \bar{Z_2}(\infty))+i\ln(\delta \bar{Z_2}(-\infty)) =
- 2i \ln(\prod_{i=1}^{N}\ \Sigma_i)\ +P, \\ \nonumber
&=& i \sum_{i=1}^N \left[ 2\ln\left( \frac{r - x_i^+}{r- x_i^-} \right)  -
\ln\left( \frac{x_i^+}{x_i^-}\right) \right].
\end{eqnarray}
The reason that the difference in phases at $\infty$ and $-\infty$
 gives the value of the phase shift at the momentum
$1/r$ is because the momentum and frequency of the complex conjugate field is of the opposite sign
compared to the field $Z_2$.

\section{Crossing equations using the antipode operation}

In this section we derive the crossing relations for the $SU(1|1)$ matrix in
(\ref{def-smat}), (\ref{def-rmat}) and  (\ref{def-rmat2})
 using  an algebraic formulation of the crossing relations in terms of
an antipode as it was done for the case of the $SU(1|2)$ matrix in ${\cal N}=4$
super Yang-Mills by \cite{Janik:2006dc}. To simplify  our notations we will introduce the
fundamental representation of $SU(1|1)$ again and some notations below.
\begin{eqnarray}
 Q|\phi\rangle = a|\psi\rangle, \qquad  S|\psi\rangle =  b |\phi\rangle, \\ \nonumber
B|\phi\rangle = (\beta+1) |\phi\rangle, \qquad B|\psi\rangle = ( \beta-1) |\psi\rangle.
\end{eqnarray}
We can write the operators of the algebra in terms of the states as
\begin{eqnarray}
\label{op-st1}
 Q = a |\psi\rangle \langle \phi|, \qquad S =  b|\phi\rangle \langle \psi |, \\ \nonumber
B = (\beta +1) |\phi\rangle \langle \phi| + ( \beta  -1) |\psi \rangle \langle\psi|.
\end{eqnarray}
We will need to take the super-transpose of these operators.
 The super-transpose  defined by
 \begin{equation}
 M^{st}_{ij} = (-1)^{d(i)d(j) + d(j) } M_{ji}.
 \end{equation}
 where $d(\phi) =0, d(\psi) =1$.
 For the generators of $SU(1|1)$ the super-transpose works out to be given by
\begin{equation}
 Q^{st} = \frac{a}{b} S, \qquad S^{st} = - \frac{b}{a} Q, \qquad B^{st} = B
\end{equation}
We also need to take the charge conjugate of these operators.  For this purpose,
let us define the charge conjugate generators as
\begin{eqnarray}
\label{op-st2}
 \bar{Q} = \bar a|\psi\rangle \langle \phi|, \qquad \bar S = \bar{b}  |\phi\rangle \langle \psi|,
\\ \nonumber
\bar B = ( \bar\beta + 1) |\phi\rangle\langle \phi| + ( \bar\beta -1) |\psi\rangle\langle \psi|.
\end{eqnarray}
For the above  charge conjugate representation,  the super-transpose is given by
\begin{equation}
  \bar{Q}^{st} = \frac{\bar a}{\bar b} \bar{S},
\qquad \bar{S}^{st} = - \frac{\bar b}{\bar a} \bar Q, \qquad \bar{B}^{st} = \bar{B}.
\end{equation}
The coefficients $\bar a, \bar c, \bar b,  \bar\beta $ must satisfy the following.
We now need to construct the charge conjugate operation. This operation must
satisfy
\begin{equation}
{\cal C} {\cal O} +\bar{\cal O}^{st} {\cal O} =0,
\end{equation}
for any operator ${\cal O}$ of the algebra.
Here we have implemented the antipode operation as  in \cite{Janik:2006dc} using
${\cal S}{\cal A}  =-{\cal A}$ where ${\cal S} $ is the antipode operation.
Writing out the above condition for the generators in $SU(1|1)$ we obtain
\begin{eqnarray}
\label{chrg-conj}
{\cal C} Q + \bar Q^{st} {\cal C} = 0 , \quad
{\cal C} S +  \bar S^{st} {\cal C} = 0, \quad
{\cal C} B + \bar B^{st} {\cal C} = 0.
\end{eqnarray}
The operator ${\cal C}$, the charge conjugation operator must be obtained from the above equations. The equations in (\ref{chrg-conj}) also
 constrain the coefficients $\bar a, \bar c, \bar b, \bar\beta$.
Substituting the form of the generators in terms of states given in
(\ref{op-st1}) and (\ref{op-st2}) we obtain
\begin{eqnarray}
{\cal C} &=& - \frac{\bar{a}}{a} |\phi\rangle \langle \psi| + |\psi\rangle\langle \phi|
= - \frac{\bar{a}}{ab}  S + \frac{1}{a} Q,
\end{eqnarray}
with the condition
\begin{equation}
a b + \bar{a}\bar{b} =0, \qquad c = -\bar c, \qquad \beta = - \bar\beta.
\end{equation}
We can also write down the inverse of $C$ as the following
\begin{eqnarray}
 C^{-1} & =& - \frac{a}{\bar{a}} |\psi\rangle\langle \phi| + |\phi\rangle \langle \psi|
= - \frac{1}{\bar{a}} Q + \frac{1}{b} S.
\end{eqnarray}

The crossing constraint  which the full $R$ matrix must satisfy according to
\cite{Janik:2006dc} is  the following
\begin{equation}
\label{janik}
 ( C^{-1} \otimes I){\cal  R}_{\bar{1} 2} (\bar\alpha_1, \alpha_2)^{st_1} ( C \otimes I)
{\cal  R}_{12} (\alpha_1, \alpha_2 ) = I.
\end{equation}
where super-transpose acts on the first Hilbert space.
From the structure of the $R$ matrix in (\ref{def-rmat}) and (\ref{def-rmat2})
 we see that we need the
 following relation to implement the crossing condition
\begin{eqnarray}
\label{conj-cas}
&&(C^{-1} \otimes I) (J_{\bar{1} 2}^2)^{st} (C\otimes I)  \\ \nonumber
&=& 2\beta_1g c_1 + 2\beta_2g c_2 - 2 B^{(1)} C^{(2)} - 2C^{(1)} B^{(2)} -
4 Q^{(1)} S^{(1)} + 4 S^{(1)} Q^{(2)}, \\ \nonumber
&=& 4 \beta_1 c_1 + 4 \beta_2c_2  - J_{12}^2.
\end{eqnarray}
Note that here we have used the equations in \eq{chrg-conj} as well as the relations
$\bar \beta_1 = -\beta_1, \bar c_1 = -\bar c_1$. To obtain the second line we
have used the definition of the Casimir given in (\ref{def-cas}).
Now using \eq{conj-cas} and \eq{def-casq} in the condition \eq{janik} we see that
the we must require that the coefficient of $J_{12}^2$ should vanish on the LHS. This leads to
the following equation
\begin{eqnarray}
&&[ R_{\bar{1} 2, 1} ( \bar\alpha_1, \alpha_2) +
( 4 b_1 gc_1 + 4 b_2g c_2) R_{\bar{1} 2; 2} (\bar\alpha_1, \alpha_2) ] R_{12, 2} (\alpha_1, \alpha_2)
\\ \nonumber
&&- R_{\bar{1} 2, 2}( \bar\alpha_1, \alpha_2) R_{12, 1} (\alpha_1, \alpha_2)
- 4R_{\bar{1}2, 2}(\bar\alpha_1, \alpha_2) R_{12, 2} (\alpha_1, \alpha_2) (b_1+ b_2) (gc_1+gc_2) =0.
\end{eqnarray}
On substituting the form for the coefficients of the $R$ matrix  given in \eq{def-rmat2} we
obtain the condition
\begin{equation}
\bar\alpha_1 =\alpha_1,
\end{equation}
The terms proportional to $g$ vanish
 due to the following identity
\begin{equation}
\label{iden-ap}
 \frac{1}{2} ( -\beta_1 + \beta_2) ( -c_1 +c_2) - (\beta_1 c_1 + \beta_2c_2)
-\frac{1}{2}( \beta_1+\beta_2) (c_1+c_2) + (\beta_1 \beta_2)(c_1+c_2) =0 .
\end{equation}
This is an important consistency check of our equations and the definition of the
conjugation operation.
Now the term in \eq{janik} which is proportional to identity leads to the following equation
\begin{eqnarray}
&& R_{\bar{1}2,0}(\alpha_1, \alpha_2) R_{12, 0}(\alpha_1, \alpha_2)  \\ \nonumber
& &\times
\frac{ (\alpha_2 -\alpha_1)^2 + \frac{1}{4}(gc_1 +gc_2)^2 }{
( (\alpha_2 -\alpha_1) - \frac{i}{2}  ( -gc_1 +g c_2) ) (  (\alpha_2- \alpha_1) - \frac{i}{2} ( gc_1+gc_2) ) }
=1.
\end{eqnarray}
One of the important consistency checks of the above equation is that the
constraint should not depend on the $\beta$'s since
these are the charges of the generator $B$ which is an outer automorphism.
An identity similar to    (\ref{iden-ap}) ensures this.
Thus the phase factor has to satisfy the following constraints
\begin{eqnarray}
 R_{12, 0} (\alpha_1, \alpha_2) R_{12, 0} ( \alpha_2, \alpha_1) &=&1, \\ \nonumber
R_{\bar{1}2, 0} (\alpha_1, \alpha_2) R_{12, 0} (\alpha_1, \alpha_2) &=& f(1, 2) , \\ \nonumber
&=& \frac{  (\alpha_2 -\alpha_1) - \frac{i}{2}  ( -gc_1 + gc_2)   }
{ (\alpha_2 -\alpha_1) + \frac{i}{2}(gc_1 +gc_2) }.
\end{eqnarray}
$\alpha$ and $c$ are related to the spectral parameters by
\begin{equation}
\label{spec-rela}
\alpha = \frac{g}{2} ( x^+ + x^-), \qquad c = -i(x^+ - x^-).
\end{equation}
Substituting these expression for $\alpha$ and $c$ in terms of the
spectral parameters we obtain
\begin{equation}
 f(x, y) = \frac{y^- - x^-}{y^+ - x^-} .
\end{equation}
For completeness we mention that carrying out a similar
analysis with the conjugation operation on the particle 2 leads to the
function
\begin{equation}
g(x, y) = \frac{y^+ - x^+}{y^+ -x^-}.
\end{equation}
Note that here the functions $f$ and $g$ satisfy the unitarity constraint
\begin{equation}
f(x, y) g(y, \bar x) = 1
\end{equation}
with $\bar x$ defined as $\bar x^+ = x^-, \bar x^- = x^+$.

The conjugation operation obtained using the antipode results in
\begin{equation}
\bar\alpha = \alpha, \qquad \bar c = - c
\end{equation}
From (\ref{spec-rela}) we see that the spectral parameters for the
conjugate particle
is therefore
\begin{equation}
\bar{x^+} = x^-, \qquad \bar{x^-} = x^+
\end{equation}
As discussed in section 5, this reverses the sign of the momentum but
does not change the sign of the energy. Thus the algebraic method of
using the antipode does not result in the required conjugation which
transforms a particle to an anti-particle.

\providecommand{\href}[2]{#2}\begingroup\raggedright\endgroup


\begin{thebibliography}{10}

\bibitem{Serban:2010sr}
D.~Serban, {\it {Integrability and the {$AdS/CFT$} correspondence}},
  \href{http://xxx.lanl.gov/abs/1003.4214}{{\tt arXiv:1003.4214}}.

\bibitem{Maldacena:1998bw}
J.~M. Maldacena and A.~Strominger, {\it {{$AdS_3$} black holes and a stringy
  exclusion principle}},  {\em JHEP} {\bf 12} (1998) 005,
  [\href{http://xxx.lanl.gov/abs/hep-th/9804085}{{\tt hep-th/9804085}}].

\bibitem{Bena:2003wd}
I.~Bena, J.~Polchinski, and R.~Roiban, {\it {Hidden symmetries of the
  {$AdS_5\times S^5$} superstring}},  {\em Phys. Rev.} {\bf D69} (2004) 046002,
  [\href{http://xxx.lanl.gov/abs/hep-th/0305116}{{\tt hep-th/0305116}}].

\bibitem{Pakman:2009ab}
A.~Pakman, L.~Rastelli, and S.~S. Razamat, {\it {Extremal Correlators and
  Hurwitz Numbers in Symmetric Product Orbifolds}},  {\em Phys. Rev.} {\bf D80}
  (2009) 086009, [\href{http://xxx.lanl.gov/abs/0905.3451}{{\tt
  arXiv:0905.3451}}].

\bibitem{Pakman:2009zz}
A.~Pakman, L.~Rastelli, and S.~S. Razamat, {\it {Diagrams for Symmetric Product
  Orbifolds}},  {\em JHEP} {\bf 10} (2009) 034,
  [\href{http://xxx.lanl.gov/abs/0905.3448}{{\tt arXiv:0905.3448}}].

\bibitem{Pakman:2009mi}
A.~Pakman, L.~Rastelli, and S.~S. Razamat, {\it {A Spin Chain for the Symmetric
  Product {$CFT_2$} }},  \href{http://xxx.lanl.gov/abs/0912.0959}{{\tt
  arXiv:0912.0959}}.

\bibitem{Babichenko:2009dk}
A.~Babichenko, B.~Stefanski, Jr., and K.~Zarembo, {\it {Integrability and the
  {$AdS_3/CFT_2$} correspondence}},  {\em JHEP} {\bf 03} (2010) 058,
  [\href{http://xxx.lanl.gov/abs/0912.1723}{{\tt arXiv:0912.1723}}].

\bibitem{David:2008yk}
J.~R. David and B.~Sahoo, {\it {Giant magnons in the D1-D5 system}},  {\em
  JHEP} {\bf 07} (2008) 033, [\href{http://xxx.lanl.gov/abs/0804.3267}{{\tt
  arXiv:0804.3267}}].

\bibitem{Chen:2007vs}
H.-Y. Chen, N.~Dorey, and R.~F. Lima~Matos, {\it {Quantum Scattering of Giant
  Magnons}},  {\em JHEP} {\bf 09} (2007) 106,
  [\href{http://xxx.lanl.gov/abs/0707.0668}{{\tt arXiv:0707.0668}}].

\bibitem{David:2002wn}
J.~R. David, G.~Mandal, and S.~R. Wadia, {\it {Microscopic formulation of black
  holes in string theory}},  {\em Phys. Rept.} {\bf 369} (2002) 549--686,
  [\href{http://xxx.lanl.gov/abs/hep-th/0203048}{{\tt hep-th/0203048}}].

\bibitem{Gava:2002xb}
E.~Gava and K.~S. Narain, {\it {Proving the {PP}-wave / CFT{${}_2$} duality}},
  {\em JHEP} {\bf 12} (2002) 023,
  [\href{http://xxx.lanl.gov/abs/hep-th/0208081}{{\tt hep-th/0208081}}].

\bibitem{Arutyunov:1997gi}
G.~E. Arutyunov and S.~A. Frolov, {\it {Four graviton scattering amplitude from
  {$S(N)R^8$} supersymmetric orbifold sigma model}},  {\em Nucl. Phys.} {\bf
  B524} (1998) 159--206, [\href{http://xxx.lanl.gov/abs/hep-th/9712061}{{\tt
  hep-th/9712061}}].

\bibitem{Jevicki:1998bm}
A.~Jevicki, M.~Mihailescu, and S.~Ramgoolam, {\it {Gravity from {CFT} on
  {$S^N(X)$}: Symmetries and interactions}},  {\em Nucl. Phys.} {\bf B577}
  (2000) 47--72, [\href{http://xxx.lanl.gov/abs/hep-th/9907144}{{\tt
  hep-th/9907144}}].

\bibitem{Lunin:2000yv}
O.~Lunin and S.~D. Mathur, {\it {Correlation functions for {$M^N/S^N$}
  orbifolds}},  {\em Commun. Math. Phys.} {\bf 219} (2001) 399--442,
  [\href{http://xxx.lanl.gov/abs/hep-th/0006196}{{\tt hep-th/0006196}}].

\bibitem{Lunin:2001pw}
O.~Lunin and S.~D. Mathur, {\it {Three-point functions for {$M^N/S(N)$}
  orbifolds with {${\cal N} = 4$} supersymmetry}},  {\em Commun. Math. Phys.}
  {\bf 227} (2002) 385--419,
  [\href{http://xxx.lanl.gov/abs/hep-th/0103169}{{\tt hep-th/0103169}}].

\bibitem{Gomis:2002qi}
J.~Gomis, L.~Motl, and A.~Strominger, {\it {{PP}-wave / CFT{${}_2$} duality}},
  {\em JHEP} {\bf 11} (2002) 016,
  [\href{http://xxx.lanl.gov/abs/hep-th/0206166}{{\tt hep-th/0206166}}].

\bibitem{Lunin:2002fw}
O.~Lunin and S.~D. Mathur, {\it {Rotating deformations of {$AdS_3\times S_3$},
  the orbifold {CFT} and strings in the pp-wave limit}},  {\em Nucl. Phys.}
  {\bf B642} (2002) 91--113,
  [\href{http://xxx.lanl.gov/abs/hep-th/0206107}{{\tt hep-th/0206107}}].

\bibitem{Hikida:2002in}
Y.~Hikida and Y.~Sugawara, {\it {Superstrings on {PP}-wave backgrounds and
  symmetric orbifolds}},  {\em JHEP} {\bf 06} (2002) 037,
  [\href{http://xxx.lanl.gov/abs/hep-th/0205200}{{\tt hep-th/0205200}}].

\bibitem{Berenstein:2002jq}
D.~E. Berenstein, J.~M. Maldacena, and H.~S. Nastase, {\it {Strings in flat
  space and pp waves from {${\cal N} = 4$} super {Y}ang {M}ills}},  {\em JHEP}
  {\bf 04} (2002) 013, [\href{http://xxx.lanl.gov/abs/hep-th/0202021}{{\tt
  hep-th/0202021}}].

\bibitem{Hofman:2006xt}
D.~M. Hofman and J.~M. Maldacena, {\it Giant magnons},  {\em J. Phys.} {\bf
  A39} (2006) 13095--13118, [\href{http://xxx.lanl.gov/abs/hep-th/0604135}{{\tt
  hep-th/0604135}}].

\bibitem{Maldacena:2006rv}
J.~M. Maldacena and I.~Swanson, {\it {Connecting giant magnons to the pp-wave:
  An interpolating limit of {$AdS_5 \times S^5$}}},  {\em Phys. Rev.} {\bf D76}
  (2007) 026002, [\href{http://xxx.lanl.gov/abs/hep-th/0612079}{{\tt
  hep-th/0612079}}].

\bibitem{Chen:2006gea}
H.-Y. Chen, N.~Dorey, and K.~Okamura, {\it {Dyonic giant magnons}},  {\em JHEP}
  {\bf 09} (2006) 024, [\href{http://xxx.lanl.gov/abs/hep-th/0605155}{{\tt
  hep-th/0605155}}].

\bibitem{Arutyunov:2006gs}
G.~Arutyunov, S.~Frolov, and M.~Zamaklar, {\it {Finite-size effects from giant
  magnons}},  {\em Nucl. Phys.} {\bf B778} (2007) 1--35,
  [\href{http://xxx.lanl.gov/abs/hep-th/0606126}{{\tt hep-th/0606126}}].

\bibitem{Spradlin:2006wk}
M.~Spradlin and A.~Volovich, {\it {Dressing the giant magnon}},  {\em JHEP}
  {\bf 10} (2006) 012, [\href{http://xxx.lanl.gov/abs/hep-th/0607009}{{\tt
  hep-th/0607009}}].


\bibitem{Minahan:2006bd}
J.~A. Minahan, A.~Tirziu, and A.~A. Tseytlin, {\it {Infinite spin limit of
  semiclassical string states}},  {\em JHEP} {\bf 08} (2006) 049,
  [\href{http://xxx.lanl.gov/abs/hep-th/0606145}{{\tt hep-th/0606145}}].

\bibitem{Beisert:2005wm}
N.~Beisert, {\it {An {$SU(1|1)$}-invariant S-matrix with dynamic
  representations}},  {\em Bulg. J. Phys.} {\bf 33S1} (2006) 371--381,
  [\href{http://xxx.lanl.gov/abs/hep-th/0511013}{{\tt hep-th/0511013}}].

\bibitem{Beisert:2004hm}
N.~Beisert, V.~Dippel, and M.~Staudacher, {\it {A novel long range spin chain
  and planar {${\cal N} = 4$} super Yang- Mills}},  {\em JHEP} {\bf 07} (2004)
  075, [\href{http://xxx.lanl.gov/abs/hep-th/0405001}{{\tt hep-th/0405001}}].

\bibitem{Jackiw:1975im}
R.~Jackiw and G.~Woo, {\it {Semiclassical Scattering of Quantized Nonlinear
  Waves}},  {\em Phys. Rev.} {\bf D12} (1975) 1643.

\bibitem{Hernandez:2006tk}
  R.~Hernandez and E.~Lopez,
  {\it { Quantum corrections to the string Bethe ansatz}},
  {\em JHEP} {\bf 0607} (2006) 004,
 [\href{http://xxx.lanl.gov/abs/hep-th/0603204}{{\tt hep-th/0603204}}].


\bibitem{Beisert:2006ib}
  N.~Beisert, R.~Hernandez and E.~Lopez,
  {\it{ A crossing-symmetric phase for $AdS_5 \times  S^5$ strings}},
  {\em JHEP}  {\bf 0611} (2006) 070,
[\href{http://xxx.lanl.gov/abs/hep-th/0609044}{{\tt hep-th/0609044}}].
 





\bibitem{Arutyunov:2004vx}
G.~Arutyunov, S.~Frolov, and M.~Staudacher, {\it {Bethe ansatz for quantum
  strings}},  {\em JHEP} {\bf 10} (2004) 016,
  [\href{http://xxx.lanl.gov/abs/hep-th/0406256}{{\tt hep-th/0406256}}].

\bibitem{Chen:2006gq}
H.-Y. Chen, N.~Dorey, and K.~Okamura, {\it {On the scattering of magnon
  boundstates}},  {\em JHEP} {\bf 11} (2006) 035,
  [\href{http://xxx.lanl.gov/abs/hep-th/0608047}{{\tt hep-th/0608047}}].

\bibitem{Kazakov:2004qf}
V.~A. Kazakov, A.~Marshakov, J.~A. Minahan, and K.~Zarembo, {\it {Classical /
  quantum integrability in AdS/CFT}},  {\em JHEP} {\bf 05} (2004) 024,
  [\href{http://xxx.lanl.gov/abs/hep-th/0402207}{{\tt hep-th/0402207}}].

\bibitem{Berkovits:1999im}
N.~Berkovits, C.~Vafa, and E.~Witten, {\it {Conformal field theory of {$AdS$}
  background with Ramond- Ramond flux}},  {\em JHEP} {\bf 03} (1999) 018,
  [\href{http://xxx.lanl.gov/abs/hep-th/9902098}{{\tt hep-th/9902098}}].

\bibitem{Janik:2006dc}
R.~A. Janik, {\it {The {$AdS_5\times S^5$} superstring worldsheet S-matrix and
  crossing symmetry}},  {\em Phys. Rev.} {\bf D73} (2006) 086006,
  [\href{http://xxx.lanl.gov/abs/hep-th/0603038}{{\tt hep-th/0603038}}].

\bibitem{Arutyunov:2006iu}
G.~Arutyunov and S.~Frolov, {\it {On {$AdS_5\times S^5$} string S-matrix}},
  {\em Phys. Lett.} {\bf B639} (2006) 378--382,
  [\href{http://xxx.lanl.gov/abs/hep-th/0604043}{{\tt hep-th/0604043}}].

\bibitem{Gomez:2006va}
  C.~Gomez and R.~Hernandez,
  {\it{ The magnon kinematics of the AdS/CFT correspondence}},
  {\em JHEP} {\bf 0611} (2006) 021,
[\href{http://xxx.lanl.gov/abs/hep-th/0608029}{{\tt hep-th/0608029}}].

\bibitem{Plefka:2006ze}
  J.~Plefka, F.~Spill and A.~Torrielli,
  {\it{ On the Hopf algebra structure of the AdS/CFT S-matrix}},
{\em Phys. Rev.} {\bf D74}
  (2006) 066008, [\href{http://xxx.lanl.gov/abs/hep-th/0608038}{{\tt
  hep-th/0608038}}].


\bibitem{Arutyunov:2006yd}
G.~Arutyunov, S.~Frolov, and M.~Zamaklar, {\it {The Zamolodchikov-Faddeev
  algebra for {$AdS_5\times S^5$} superstring}},  {\em JHEP} {\bf 04} (2007)
  002, [\href{http://xxx.lanl.gov/abs/hep-th/0612229}{{\tt hep-th/0612229}}].

\bibitem{Kalousios:2008gz}
  C.~Kalousios, G.~Papathanasiou and A.~Volovich,
  {\it {Exact solutions for $N$-magnon scattering}},
  {\em JHEP } {\bf 0808} (2008) 095
    [\href{http://xxx.lanl.gov/abs/0806.2466}{{\tt arXiv:0806.2466}}].


\end{thebibliography}

\end{document}